\documentclass[10pt,journal,compsoc]{IEEEtran}

\makeatletter
\def\ps@headings{%
	\def\@oddhead{\mbox{}\scriptsize\rightmark \hfil \thepage}%
	\def\@evenhead{\scriptsize\thepage \hfil \leftmark\mbox{}}%
	\def\@oddfoot{}%
	\def\@evenfoot{}}
\makeatother
\usepackage{amsmath}

\usepackage{epsfig}
\usepackage{amssymb}
\usepackage{tabularx}
\usepackage[skip=1ex]{caption}
\usepackage{multirow}
\usepackage{rotating}
\usepackage{graphicx}
\usepackage{tabularx}
\usepackage{epstopdf}
\usepackage{array}
\usepackage{mathrsfs}
\usepackage{amsthm} 
\usepackage{amsfonts}
\usepackage{multicol}
\usepackage{epsfig}
\usepackage{algpseudocode}
\usepackage{csquotes}
\usepackage{tabu}
\usepackage{float}
\usepackage{url}
\usepackage{blindtext}
\usepackage{cite}
\usepackage{lipsum}
\usepackage{mathtools}
\usepackage{cuted}
\usepackage{etoolbox}
\usepackage{flushend}
\usepackage{subcaption}
\usepackage{stackengine}
\usepackage{hyperref}
\newcommand\xrowht[2][0]{\addstackgap[.5\dimexpr#2\relax]{\vphantom{#1}}}
\usepackage[ruled,vlined,linesnumbered,resetcount]{algorithm2e}
\SetKwRepeat{Do}{do}{while}%
\newtheorem*{thm1}{Theorem~1}

\newtheorem*{propos1}{Proposition~1}

\usepackage[linesnumbered,ruled,vlined]{algorithm2e}
\SetKwRepeat{Do}{do}{while}%
\usepackage{url}

\usepackage[latin1]{inputenc}
\newcounter{BlockCounter}
\usepackage{tikz}
\usetikzlibrary{shapes,arrows,positioning}
\makeatletter
\renewcommand{\SetKwInOut}[2]{%
	\sbox\algocf@inoutbox{\KwSty{#2}\algocf@typo:}%
	\expandafter\ifx\csname InOutSizeDefined\endcsname\relax% if first time used
	\newcommand\InOutSizeDefined{}%
	\sbox\algocf@inoutbox{\KwSty{#2}\algocf@typo\textbf{:}~}\setlength{\inoutindent}{\wd\algocf@inoutbox}%
	\else% else keep the larger dimension
	\ifdim\wd\algocf@inoutbox>\inoutsize%
	\sbox\algocf@inoutbox{\KwSty{#2}\algocf@typo\textbf{:}~}\setlength{\inoutindent}{\wd\algocf@inoutbox}%
	\fi%
	\fi% the dimension of the box is now defined.
	\algocf@newcommand{#1}[1]{%
		\ifthenelse{\boolean{algocf@inoutnumbered}}{\relax}{\everypar={\relax}}%
		{\let\\\algocf@newinout\hangindent=\inoutindent\hangafter=1\KwSty{#2}\algocf@typo\textbf{:}~##1\par}%
		\algocf@linesnumbered% reset the numbering of the lines
}}%

\makeatother

\begin{document}
	%%%%%%%%%%%%%%%%%%%%%%%%%%%%%%%%%%%%%%%%%%%%%%%%%%%%%%
	\newcommand{\labelBlock}[1]{%
		\refstepcounter{BlockCounter}%
		\hypertarget{#1}{}(\theBlockCounter\label{#1})%
	}
	
	\newcommand{\refBlock}[1]{%
		\hyperref[#1]{Block~\ref*{#1}}% (see Problem 18 of the hyperref manual)
	}
	\newcommand*\colvec[1]{\begin{pmatrix}#1\end{pmatrix}}
	\makeatletter
	\newcommand*{\rom}[1]{\expandafter\@slowromancap\romannumeral #1@}
	\makeatother
	
	\title{Energy and Cost Efficient Resource Allocation for Blockchain-Enabled NFV}% ,~\IEEEmembership{Student Member,~IEEE}  ,
	\author{Shiva Kazemi Taskou, Mehdi Rasti,~\IEEEmembership{Member,~IEEE}, and  Pedro H. J. Nardelli,~\IEEEmembership{Senior Member,~IEEE} 
		\thanks{S. Kazemi Taskou and M. Rasti are with Department of Computer Engineering, Amirkabir University of Technology, Tehran, Iran. (e-mail: \{shiva.kt,  rasti\}@aut.ac.ir). M. Rasti is also a visiting assitant professor in Lappeenranta-Lahti University of Technology, Lappeenranta, Finland.}
		\thanks{Pedro H. J. Nardelli is with  Lappeenranta-Lahti University of Technology, Lappeenranta, Finland. (e-mail:Pedro.Nardelli@lut.fi)}
		\thanks{This work is supported by the Academy of Finland via: (a) ee-IoT project n.319009, and, (b) EnergyNet Research Fellowship n.321265/n.328869 and (c) FIREMAN consortium n.326270 as part of CHIST-ERA grant CHIST-ERA-17-BDSI-003. }
	}
	%\maketitle
	
	%\markboth{Copyright (c) 2017 IEEE. Personal use of this material is permitted. However, permission to use this material for any other purposes must be obtained from the IEEE by sending a request to pubspermissions@ieee.org.}	{Kazemi Taskou \MakeLowercase{\textit{et al.}}: Fast Water-Filling Method for Sum-Power Minimization in OFDMA  Networks}

	%\IEEEpeerreviewmaketitle
	\IEEEtitleabstractindextext{%
		
		%=======================   Abstract =========================		
		\begin{abstract}\
			Network function virtualization (NFV) is a promising technology to make  5G networks flexible and agile. NFV  decreases operators' OPEX and CAPEX  by decoupling the physical hardware from the functions they perform. In NFV, users' service request can be viewed as a service function chain (SFC) consisting of several virtual network functions (VNFs) which are connected through virtual links.  
			Resource allocation  in  NFV is done through a centralized authority called NFV Orchestrator (NFVO). This centralized authority suffers from some drawbacks such as single point of failure and security. Blockchain (BC) technology is able to address these problems by decentralizing resource allocation.
			The drawbacks of NFVO in NFV architecture and the exceptional BC characteristics to address these problems motivate us to focus on NFV resource allocation to users'  SFCs without the need for an NFVO. To this end,  we assume there are two types of users: users who send SFC requests (SFC requesting users) and users who perform mining process (miner users). For SFC requesting users, we formulate  NFV resource allocation (NFV-RA)  problem as a multi-objective problem to minimize the energy consumption and utilized resource cost, simultaneously. To address this problem, we propose an Approximation-based Resource Allocation algorithm (ARA) using Majorization-Minimization approximation method to convexify NFV-RA problem. Furthermore, due to the high complexity of ARA algorithm, we propose a low complexity Hungarian-based Resource Allocation (HuRA) algorithm using Hungarian algorithm for server allocation. Through the simulation results, we show that our proposed  ARA and HuRA algorithms  achieve near-optimal performance  with lower computational complexity. Also,  ARA algorithm  outperforms the existing algorithms in terms of number of active servers, energy consumption, and average latency.
			Moreover, the mining process is the foundation of BC technology. In wireless networks, mining is performed by resource-limited mobile users. Since the mining process requires high computational complexity, miner users cannot perform it alone. So, in this paper, we assume that miner users can perform mining process with participating of other users. For mining process, the problem of minimizing the energy consumption and  cost of  users' processing resources  is formulated as a linear programming problem that can be optimally solved in polynomial time.  
		\end{abstract}
		
		}	\maketitle

	\IEEEdisplaynontitleabstractindextext
	
	\IEEEpeerreviewmaketitle
	%=======================    Introduction  ========================
	\vspace{-1.5 em}
	\IEEEraisesectionheading{\section{Introduction}}
	With the ever-increasing users' traffic and demands on new services in 5G networks, flexibility, scalability, and agility are inevitable. To meet these requirements,  Network Function Virtualization (NFV)  has attracted great attention from both industry and academia \cite{access-2019-intro}. The main idea of NFV  defined by the European Telecom Standards Institute (ETSI) is decoupling the physical infrastructures from the functions running on them \cite{survey-NFV}. With NFV, network functions that are traditionally run on dedicated hardware which results in high CAPEX and OPEX are typically implemented as Virtual Network Functions (VNFs) on commodity devices in data centers \cite{survey-TNSM}.

	In NFV, the users' service requests called Service Function Chains (SFCs) consist of several different VNFs interconnected by virtual links in a given order. To serve SFCs, VNFs should be implemented on commodity devices and the processing resources should be allocated to them, and the communication between them must be provided by physical links bandwidth allocation such that the users' requirements are satisfied \cite{access-2019-intro-2}. One of the main challenges of NFV is the allocation of processing resources  (known as VNF placement) and physical links bandwidth allocation to communicate between VNFs (known as routing) to execute SFCs \cite{NFV-access-2018}.
	
	According to the NFV architecture provided by ETSI, NFV Orchestrator (NFVO) is responsible for the creation and life cycle management of SFCs, management of VNFs, NFV infrastructure, and SFCs \cite{comst-dependability}. Furthermore, NFVO collects the service requests and performs a resource allocation algorithm to allocate processing resources and physical links bandwidth to the SFCs \cite{survey-TNSM,comst-dependability}. This architecture relying on  NFVO suffers from several drawbacks:  (1) vulnerability to failure and the outage due to centralized management of NFV infrastructure %(i.e., if the NFVO is unavailable, the system will not be able to provide any services) %(3) correct-less possibility due to timing error 
	\cite{comst-dependability},  (2) the high probability and impact of the attack due to the shared NFV infrastructure between the tenants, and (3) the need for trust-based resource allocation and management in an NFV environment that is inherently trust-less \cite{NFV-BC-ICC-2019}.   Recently, blockchain  technology has  been regarded as a promising decentralized technology to  address these pitfalls. 
	
	Blockchain (BC) is a decentralized ledger in which trusted data is stored in an untrusted environment in transaction format. In BC, all users and nodes of the network can communicate over a point-to-point network without the need of a centralized trusted entity \cite{nist}. In the second generation of BC networks known as Ethereum, smart contract-based BC was proposed \cite{survey-SC}. Smart contracts are computer programs that facilitate contracts between two or more parties. The smart contracts store all the rules agreed by the parties. These smart contracts are stored in the BC after verifying by all parties. The immutability of smart contracts after storing on the BC provides security for parties \cite{survey-SC}. In smart contract-based BC, only transactions that satisfy all the conditions of stored smart contracts are validated \cite{survey-SC}. 
	
	Verification of smart contracts and validation of transactions and their ordering in blocks are done based on a consensus mechanism. In the consensus mechanism, nodes and parties agree on the BC status, which results in a single BC throughout the whole BC network verifying by all nodes, although this BC is stored in distributed nodes \cite{consensus-access}. Some generations of BC networks such as Bitcoin \footnote{Bitcoin is the first generation of blockchain to offer bitcoin exchange, a kind of cryptocurrency. Interested readers refer to \cite{bitcoin} for further explanation.}  use an incentive-based consensus named   mining process in their protocol to reach agreement in an untrusted environment and among a large number of distributed nodes \cite{consensus-access}. Through the mining process, the nodes that perform the mining named miners insert a number of validated transactions into a new block, which is uniquely identified by its hash and time-stamp. In Bitcoin, miners employ the Proof-of-Work (PoW) algorithm for mining. In PoW, miners try to find a random value called nonce for achieving the block's hash using their computation power. This nonce which should be less than a target value is set to avoid any conflicts and provide trust between nodes. The miner who finds the nonce faster than other miners wins the mining process and receives rewards from the BC network \cite{BC-survey-2018}. After generating the new block, the winning miner broadcasts it to all other nodes for verification of the block. The other nodes apply the transactions of this block and add its hash into the next block header. Each new block has the previous block's hash in its header and a chain of continuous blocks forms the BC \cite{bc-access-2018}. 
	The main characteristics of BC include decentralization, transparency, immutability, availability, and security \cite{BC-survey-2018} make it suitable for applying to NFV.
	
	On the other hand, energy consumption in data centers has many environmental, economic, and performance impacts. In addition, recent studies claim that the energy consumption of data centers worldwide in 2012 was 270 TWh \cite{energy-comst}. Therefore, reducing the energy consumption of data centers is a very important challenge \cite{energy-comst}. Furthermore, NFV has been proposed to reduce operators' costs, so reducing operators' consumed resource costs is an important challenge \cite{cost-tnsm}.
	
	The drawbacks of the NFV architecture and the exceptional BC characteristics to address these problems as well as the importance of minimizing the energy consumption and costs motivate us to focus on resource allocation to users' requested SFCs without the need for an NFVO, based on BC technology. To this end, we define the NFV Resource Allocation (NFV-RA) problem with the aim of minimizing the energy consumption and cost of utilized data center resources satisfying users' maximum end-to-end tolerable delay. Furthermore, to implement NFVO in a distributed manner, we propose a BC framework for NFV resource allocation named  NFVChain. To do so, we assume there exist two types of users in our considered system model: (1) users who request SFCs (SFC requesting users)  and (2) users who perform the mining process (miner users). For SFC requesting users, we define the NFV-RA problem as explained above. Moreover, since miner users are unable to perform the mining lonely because of their limited processing capacity, we assume each miner user performs the mining process with the help of a group of users. The miner should pay for consuming other users' processing capacity.  So, in this paper, for miner users, the problem of offloading the mining task to a group of users is defined to minimize the energy consumption in the mining process and minimize the cost of other users' processing capacity. It is worth mentioning that we call this problem as Mining Offloading  (MO) problem.
	\vspace{-0.5 em}
	\subsection{Related Works}
	In this section, we first discuss the related works from two aspects, namely resource allocation in NFV relied on ETSI proposed architecture followed by a discussion on the integration of BC technology with 5G networks.
	\vspace{-0.8 em}
	\subsubsection{Resource Allocation in NFV Relied on ETSI Proposed Architecture}
	The NFV-RA problem (i.e., VNF placement and routing) has been extensively investigated in literature \cite{NFV-access-2018}, \cite{rw-nfv-1}--\cite{rw-nfv-14}. The  problem of minimizing the end-to-end delay is considered in \cite{rw-nfv-1}--\cite{rw-nfv-3}. In \cite{rw-nfv-1}, the end-to-end delay is defined as the waiting delay in the server processing queue, while in  \cite{rw-nfv-2} and \cite{rw-nfv-3}, the end-to-end delay is defined as the transmission delay over physical links. %Additionally, the computation capacity of servers and the bandwidth of physical links is assumed limited. 
	%In\cite{rw-nfv-1}, a heuristic algorithm is proposed to address the defined optimization problem.  
	The problem of minimizing the transmission delay in \cite{rw-nfv-2} is defined as a mixed-integer linear programming (MILP) problem. To tackle the NP-hardness of this problem, a heuristic algorithm is proposed. The NFV-RA problem in \cite{rw-nfv-3} is defined as  a multi-objective problem that aims to minimize the transmission delay and link bandwidth consumption and maximize servers load rate, simultaneously.  Authors in \cite{rw-nfv-3}  proposed a heuristic algorithm based on the breadth-first-search method to solve this problem.
	The authors in \cite{rw-nfv-4}--\cite{rw-nfv-6} have proposed heuristic algorithms to minimize the number of utilized servers to reduce the cost of consumed computing resources of data centers. In \cite{rw-nfv-4} and \cite{rw-nfv-5},  two active and inactive states for each server in the data center are assumed, and the aim of the NFV-RA problem is to minimize the number of active servers. In \cite{rw-nfv-5} and \cite{rw-nfv-6}, the maximum end-to-end tolerable delay which is defined as the sum of processing delays on servers and transmission delays on links is guaranteed for each SFC request.
	The problem of minimizing the cost of utilized servers and physical links is investigated in \cite{NFV-access-2018}, \cite{rw-nfv-7}--\cite{rw-nfv-10}. In  \cite{rw-nfv-7}, this problem is formulated as an ILP problem and an online algorithm has been proposed to address it. %Additionally, in \cite{rw-nfv-7} for each virtual link between VNFs in SFCs, a maximum delay is considered. 
	The NFV-RA problem in \cite{NFV-access-2018} is defined as a multi-objective problem with the aim of minimizing the cost of utilized servers and links and the cost of transmission delay over links. A heuristic algorithm is proposed to solve this problem in \cite{NFV-access-2018}.
	In \cite{rw-nfv-11}--\cite{rw-nfv-12}, the servers and physical links are allocated to SFCs such that as many as possible SFC requests are admitted. Authors in \cite{rw-nfv-11} proposed an online heuristic algorithm to maximize the number of admitted SFCs satisfying the end-to-end transmission delay requirement. In \cite{rw-nfv-12}, the NFV-RA problem is formulated as a multi-objective problem to maximize the number of admitted SFCs and minimize the amount of utilized resources, simultaneously. %In \cite{rw-nfv-12}, a heuristic algorithm is proposed to address this problem.
	The authors in \cite{rw-nfv-13}  proposed a heuristic algorithm to address the NFV-RA problem which is defined as a multi-objective problem with the aim of minimizing the utilized links and maximizing servers' utilization.
	The problem of minimizing the amount of required resources  to accept all SFC requests is addressed in \cite{rw-nfv-14}. %The constraints of the servers' computational capacity and the physical links bandwidth are considered in \cite{rw-nfv-13} and \cite{rw-nfv-14}. 
	Furthermore, in \cite{rw-nfv-14}, for each SFC request, the end-to-end delay  which is defined as the sum of the transmission delay on the links is satisfied.

	\subsubsection{Integration of BC Technology With 5G Networks}
	The application of BC technology in 5G networks is investigated in several existing works \cite{rw-bc-1}--\cite{rw-bc-7}. In \cite{rw-bc-1}, BC technology is used to solve the problem of selfish behavior of relays in cooperative networks. In addition, traditionally,  the trading between relays and users is modeled as an auction algorithm. It is worth noting that the auction algorithm is relied on third-party and suffers from drawbacks such as privacy, trust, and a single point of failure. To tackle these problems in \cite{rw-bc-1}, communication between users and relays is provided through smart contract-based blockchain. %The power control problem aimed at minimizing the sum of the transmitted power of the relays and the jamming interference subject to users' required capacity is defined for relays in \cite{rw-bc-1}. This problem is transformed into a geometric programming problem that is solved by the CVX tool. 
	In the BC framework proposed in \cite{rw-bc-1}, users send their requests to relays by calling smart contract functions. Then, relays solve a power control optimization problem to calculate the best transmit power needed to send users requests. The relays then announce the cost that users have to pay for the transmission power by calling functions in the smart contract. Finally, users and relays exchange payment by calling smart contract functions. 
	In \cite{rw-bc-2}--\cite{rw-bc-3}, BC technology is implemented to enable communication between users and Mobile Edge Computing (MEC) servers which is traditionally established in a centralized manner. In \cite{rw-bc-2}, a BC-based framework is proposed in which users submit their requests to BC, other users and MEC servers decide about MEC server allocation to the users' requests, performing a matching algorithm.
	Users' requests and  MEC servers' responses are stored in BC history after the mining process.
	In \cite{rw-bc-3}, MEC servers announce their available resources by calling smart contract functions. Then, users decide to execute their tasks locally or offloading to the MEC servers by solving an optimization problem aiming at processing cost minimization. If they decide to offload, they request their required computation resources by calling functions in the smart contract. Based on the functions of this smart contract, users pay the costs to the MEC servers and the MEC servers lease their resources to the users.
	%In  \cite{rw-bc-4}, BC is employed as a decentralized data sharing method for video streaming that is traditionally performed based on centralized schemes.
	The authors in \cite{rw-bc-5}--\cite{rw-bc-6} propose BC-based approaches for virtual wireless networks. In virtual wireless networks, there is a broker that leases resources from the infrastructure provider (InP)  and rents them to the mobile virtual network operators (MVNOs). This centralized broker suffers from a single point of failure and should be trusted. To overcome these problems, BC is a promising technology to implement a broker. In the proposed frameworks in \cite{rw-bc-5}--\cite{rw-bc-6},  MVNOs send their resource requirements in a transaction and the InP announces the cost of their resources in response to this transaction. These transactions are added to the blocks and after mining stored in BC history.	
	 In \cite{rw-bc-7}, a permissioned BC is employed to reach a consensus to securely collect and synchronize information among multiple NFV management and orchestration (MANO) systems, where the required computation of the BC network is assumed to provided by MEC servers. Furthermore, the allocation of MEC servers' computational resources to the BC network is done to minimize users' cost and improve BC throughput.  
	
	To the best of our knowledge, there is no work in the literature that studies the  NFV-RA problem considering BC technology. In addition, the problem of minimization of the energy consumption of servers and the cost of utilized resources for NFV-RA  is addressed for the first time in this paper. To alleviate these drawbacks, in this paper, we focus on the blockchain-enabled NFV-RA. To do so,  the problem of minimizing the servers' energy consumption and minimizing the cost of utilized resources is formally stated as a multi-objective problem. Similar to \cite{NFV-access-2018}, \cite{rw-nfv-1}--\cite{rw-nfv-14}, the processing capacity of servers and the bandwidth of physical links are assumed to be limited. In addition, similar to \cite{rw-nfv-1}--\cite{rw-nfv-3}, \cite{rw-nfv-6}--\cite{rw-nfv-7}, \cite{rw-nfv-10}, and \cite{rw-nfv-14}, the end-to-end tolerable delay for each SFC request is guaranteed. Despite \cite{rw-nfv-2}--\cite{rw-nfv-3}, \cite{rw-nfv-10}, and \cite{rw-nfv-14} in which only the transmission delay over links is considered as the end-to-end delay, in this paper, not only the transmission delay on physical links but also the processing delay in servers are considered which is more practical. The authors in \cite{NFV-access-2018}, \cite{rw-nfv-1}--\cite{rw-nfv-14} propose algorithms to allocate resources  to the SFC requests through NFVO which is a centralized authority, in this paper, resource allocation to  SFCs is performed through a smart contract-based BC as a decentralized approach in comparison with \cite{NFV-access-2018}, \cite{rw-nfv-1}--\cite{rw-nfv-14}. 
	 Moreover, in contrast to \cite{rw-bc-7} in which a permissioned BC is employed to reach consensus among multiple MANOs, we employ a more practical permissionless blockchain in which any node can act as a miner, and there is no need to trust in miners. Besides, in \cite{rw-bc-7} the computing resources for performing the mining process are provided by MEC servers, while in this paper, we assume that the mining process is done by participating a group of users.  	
	 Furthermore, the mining process to generate a new block is performed by resource-limited mobile devices that will not be able to perform the mining process lonely due to the high computational complexity of the mining process. Hence, in our system model, the user who wants to perform the mining process (i.e., miner user) performs the mining process in collaboration with a group of other users. In this way, the miner user should pay to the other users for consuming their processing capacity. Therefore,  for miner users, the problem of mining task offloading to a group of users is defined in order to minimize the miners' energy consumption and the cost of other users' processing capacity. 
	 \vspace{-0.5 em}
	\subsection{Our Contribution}
	Our main contributions are described as follows:
	\begin{itemize}
		\item In this paper, we exploit the advantages of BC technology to overcome the problems of NFV-RA through NFVO and propose a trusted, decentralized, and secure framework so-called NFVChain. In NFVChain, we consider two types of users, namely, users who request SFCs (SFC requesting users) and users who perform the mining process as miner users. In fact, in NFVChain, the role of NFVO  is distributed among miner users.
		
		%\item  In this paper, we propose a blockchain framework for NFV resource allocation named as NFVChain. In NFVChain, the InP first states its available resources by calling the corresponding function in the smart contract. Each user then notifies its requirement and information by calling the corresponding function in the smart contract. InP collects all users' requests and then performs resource allocation algorithms and sends the results to each user. Finally, the users pay the cost of the resources.  Miners should verify these transactions and generate a new block and the winning miner adds its block to the BC.
		
		\item In the system model, we consider two types of users, users who send SFC requests  (SFC requesting users) and users who perform mining process (miner users). 
		\begin{itemize}
			\item	 For SFC requesting users, the NFV-RA problem aimed at minimizing the servers' energy consumption and the cost of utilized processing resources and bandwidth of  links is defined as a multi-objective problem satisfying the end-to-end tolerable delay for each SFC. This problem is a MILP problem, and due to its NP-hardness, we convert binary variables into continuous variables using a penalty function. Then we transform the problem into an LP problem using the Majorization-Minimization approximation method which can be solved by off-the-shelf optimization software packages. We call this algorithm as \textbf{ARA} (Approximation-based Resource Allocation). Additionally, to reduce the complexity of  ARA algorithm, we propose a heuristic algorithm based on the Hungarian method named \textbf{HuRA} (Hungarian-based Resource Allocation).
			
			\item For miner users, since they are unable to do the mining lonely because of their limited processing capacity, each miner performs the mining process with the help of a group of users. The miner user should pay for consuming other users' processing capacity. A miner user who completes the mining process faster than other miners adds a new block to the BC and receives rewards. In this paper, for miner users, the problem of offloading the mining task to a group of users (i.e., MO problem) is defined to minimize the energy consumption in the mining process and minimize the cost of other users' processing capacity.   This problem is an LP problem and can be optimally solved by off-the-shelf optimization software packages in polynomial time.	
		\end{itemize}
		\item The simulation results demonstrate that our proposed ARA and HuRA algorithms to address the NFV-RA problem obtain near-optimal performance with very low complexity. Moreover, the performance of the optimal solution of the MO problem is shown via simulation results.
	\end{itemize}
	
	The remainder of this paper is organized as follows. In Section \ref{system model}, we introduce the system model and notations. In Section \ref{problem statement}, we formally state the NFV-RA problem. The  MO problem for mining process and its solution are proposed in Section \ref{mining}. The proposed algorithms to address the stated NFV-RA problem is presented in Section \ref{ARA HuRA}.  Finally, the simulation results and conclusion are presented in Section \ref{simulation} and Section \ref{conclusion}, respectively.  
	\vspace{-0.75 em} 
	\section{System Model and Notations}\label{system model}
	Consider a BC-enabled NFV network which consists of two types of users: SFC requesting users   and miner users. Therefore, there exists a set of $ \mathcal{U} = \mathcal{U_S} \cup \mathcal{U_M}$  users, where $ \mathcal{U_S} $ and $ \mathcal{U_M} $ denote the set of SFC requesting users   and miner users, respectively. In this section, we describe the notations for NFV, framework of BC-enabled NFV (NFVChain), and notations for blockchain network, respectively. 
	 The list of  notations for NFV and blockchain network are represented, respectively, in Table. 4 and Table. 5 in Appendix A. 
	\vspace{-0.75 em} 
	\subsection{Network Function Virtualization Notations} 
	The 5G core networks consist of two types of network functions (NFs) including control plane NFs and data plane NFs. The control plane NFs such as Access and Mobility Management Function (AMF), Session Management Function (SMF) and etc., handle control plane only and data plane NF i.e., User plane function (UPF) handles data plane only \cite{3gpp}. In 5G core networks, users for receiving their desired services should connect to the network.  To do so, AMF, SMF, and UPF should be performed in order for establishing a tunnel between the users' serving base stations and the UPF.    These three NFs are required for all services, although other NFs may be implemented between these functions which their existence and order depend on the requested services \cite{3gpp}. %Recently, the implementation of the 5G core networks with NFV technology has become attractive. 
	In the NFV-based 5G core, each of the NFs is performed on commodity servers in the data center and communication between them is provided through physical links. Although in this paper, we consider 5G core virtualization, without loss of generality, this system model is applicable to virtualization of any kind of networks such as 5G radio access networks, LTE core and radio access networks,  transport networks, and the Internet.
	
	To implement an NFV-based 5G core, we assume there is a data center which is modeled as a directed graph $ \mathcal{G} = (\mathcal{V},\mathcal{L}) $, where $ \mathcal{V} $ is the set of servers and $ \mathcal{L} $ is the set of directed links. The server set $ \mathcal{V} $ can be further categorized into three disjoint subsets, i.e., $ \mathcal{V}=\{\mathcal{AC},\mathcal{TR},\mathcal{N}\} $ with $ \mathcal{AC} $ as the access switches (source nodes), $ \mathcal{TR} $ as the transport switches (destination nodes), and $ \mathcal{N} $
	as the processing servers. Each processing server $ n\in\mathcal{N} $ has a maximum processing capacity, denoted by $ C_n^{\mathrm{max}}$ CPU cycles per second. Also, the maximum traffic which can be carried by link $ l\in\mathcal{L} $  is limited to $ B_l^{\mathrm{max}} $ bits per second. It should be noticed that the access and transport switches do not have any computation capability and only forward traffic. Therefore, no limitation on processing capacity is considered for these switches.
	
	We assume there is a set of SFC requesting users denoted by $ \mathcal{U_S} $. Each SFC   consists of a number of VNFs that run sequentially. Let $ \mathcal{S}_i=\{1,2,\cdots,J_i\} $ denote the SFC for user $ i $, where $ \mathcal{S}_i[1] = $ AMF and $ \mathcal{S}_i[J_i] = $ UPF. Furthermore, $ \mathcal{S}_i[2], \mathcal{S}_i[3],\cdots , \mathcal{S}_i[J_i-1] $ contain SMF and other VNFs whose existence and order depend on the provided service type by SFC $ \mathcal{S}_i $. Additionally, by default each SFC $ \mathcal{S}_i $ has a specific source node and destination node denoted by $ \mathcal{S}_i[0] $  and $ \mathcal{S}_i[J_i +1] $, respectively. Specifically, the access switches and transport switches are considered as source and destination nodes, respectively i.e., $ \mathcal{S}_i[0]\in\mathcal{AC} $  and $ \mathcal{S}_i[J_i +1]\in\mathcal{TR} $. 
	
	To describe the embedding of the SFCs on data center, we define the binary variable $ x^{i,j}_n $ to indicate the embedding of $ j $th VNF of user $ i $'s SFC on server $ n $. If server $ n $ is chosen to perform $ j $th VNF of user $ i $'s SFC, $ x^{i,j}_n =1$, otherwise, $ x^{i,j}_n =0$. Also, $ y_{l_j^{j+1}}^i $ is a continuous variable that represents the bandwidth allocation of physical link $ l $ to virtual link between $ j $th and $ j + 1 $th VNFs of user $ i $'s SFC. Each link $ l\in\mathcal{L} $ is assumed to be bi-directional. 
	We assume that the required CPU cycles for each VNF $ j $ of SFC $ \mathcal{S}_i $ and the required bandwidth for links between $ j $th and $ j+1 $th VNFs of $ \mathcal{S}_i $ are denoted by $ C_{i,j} $  and $ B^{j,j+1}_i $, respectively.

	  It should be noticed that different SFCs require different types of VNFs, that must be performed in order, and each of these VNFs requires processing resources, which is defined as the number of required CPU cycles per second.  For example, a VNF to perform video encoding requires $ 500~ \mathrm{ MHz} $ CPU processing capacity to encode a video stream with a speed of  $ 100  $ Megabytes per second  \cite{tnsm-2016-ra-nfv}. 
	The number, type, and order of VNFs, and the processing resources required for each VNF for SFCs are determined based on the service requested in the service level agreement contract, which is available to all users \cite{etsi-sla}.    
	Furthermore, each SFC $ \mathcal{S}_i $ has an end-to-end  maximum tolerable delay requirement denoted by $ T_i^{\mathrm{th}} $. The end-to-end delay of SFC $ \mathcal{S}_i $  represented by $ T_i $ is defined as summation of processing delay on servers and transmission delay on physical links. Assuming the allocation of servers and links to SFCs according to the statistical multiplexing \cite{statistical}, the processing delay on server $ n $ and transmission delay on link $ l $ are obtained by $ {C_{i,j}}/{C_n^{\mathrm{max}}} $ and ${ y_{l_j^{j+1}}^i}/{B_l^{\mathrm{max}}}  $, respectively. Accordingly, the end-to-end delay of SFC $ \mathcal{S}_i $ is expressed as
	\begin{equation}\label{delay}
	\begin{aligned}
	T_i=\sum\limits_{n \in \mathcal{N}}\sum\limits_{j \in \mathcal{S}_i}x_n^{i,j}\dfrac{C_{i,j}}{C_n^{\mathrm{max}}} +\sum\limits_{l \in \mathcal{L}} \sum\limits_{j \in \mathcal{S}_i\cup\{0\}} \dfrac{ y_{l_j^{j+1}}^i}{B_l^{\mathrm{max}}}.
	\end{aligned}
	\end{equation}
	
	In this paper, to reduce the energy consumption in the data center, similar to \cite{rw-nfv-4}--\cite{rw-nfv-5}, we consider two active and inactive modes for servers. To represent  the mode of each server $n$, we define the binary variable $ \beta_n $; if the server  $n$ is active, $ \beta_n =1$, otherwise, $ \beta_n =0$. If at least one VNF is assigned to the server $ n $, the server will be active, otherwise, the server will be inactive. In the inactive mode, the server is off and consumes no power \cite{idle mode}. While in the active mode, the server consumes the static power $ p_n^s $ to be active and power $ p_n $ to process every CPU cycle.
	\vspace{-0.75 em}  
	\subsection{Framework of Blockchain-Enabled NFV (NFVChain)}
	To implement NFVChain, we design  a smart contract which contains all agreed  important rules between users and InP.    Each smart contract  has an address and smart contracts are triggered by transactions. Briefly, the main events that are happening in NFVChain are: (1) the smart contract is created and published on the BC, (2) the InP advertises all available resources and unit price of its resources, (3) users submit their SFC requests, (4) the InP performs the resource allocation algorithms  and sends the allocated resources and costs to the users, and (5) users pay to InP. 
	%\begin{itemize}
	%	\item The smart contract is created and published on the BC. 
	%	\item The InP advertises all available resources and unit price of its resources. 
	%	\item Users submit their SFC requests. 
	%	\item The InP performs the resource allocation algorithms  and sends the allocated resources and costs to the users. 
	%	\item Users pay to InP. 
	%\end{itemize}
	The smart contract functions for NFV-RA is illustrated in Table \ref{smart contract}. These functions are explained in more detail in what follows.
	
	\begin{table}
		\centering
		\caption{smart resource allocation contract }
		\label{smart contract}
		
		\begin{tabular}{{| m{2.5cm} | m{5cm} |}}
			\hline
			& \\
			InP-Information():  & $ \mathrm{WA_{ InP}} $, resources, cost, SFCs$ =[] $\\
			 & \\
			\hline
			%Create(): & check($ W_{\mathrm{ InP}} $, resources, cost)\\
			%&\\
			RA-Request(): & check($ \mathrm{WA_{ InP}} $, resources, cost), $ ID_{\mathrm{ SFC}} $, $ \mathrm{WA_{ SFC}} $, requirement, information,  add(SFCs, $ ID_{\mathrm{ SFC}} $)\\
			\hline
			ARA(): & verify($ ID_{\mathrm{ SFC}} $, $ \mathrm{WA_{ SFC}} $),  run(ARA algorithm), send($ ID_{\mathrm{ SFC}} $,RA results), send ($ ID_{\mathrm{ SFC}} $,cost)\\
			\hline
			HuRA():  & verify($ ID_{\mathrm{ SFC}} $, $ \mathrm{WA_{ SFC}} $), run(HuRA algorithm), send($ ID_{\mathrm{ SFC}} $,RA results), send ($ ID_{\mathrm{ SFC}} $,cost)\\
			\hline
			Payment(): & check(requirement satisfying), check(cost), send($ \mathrm{WA_{ InP}} $, cost)\\
			
			\hline
		\end{tabular}
		\vspace{-1.5 em}
	\end{table}

	\textbf{InP-Information()}. By calling this function, the InP advertises its wallet address ($ \mathrm{WA_{ InP}} $), the unit price of its resources (cost), the available resources,  and the information required by  users including the power consumption of  servers, the maximum processing capacity of  servers, and maximum bandwidth of physical links (resources). Furthermore, it creates an empty set of SFCs (SFCs$ =[] $). InP then signs this transaction with its private key. Then it broadcasts the signed transaction to all users. By doing so, users are informed that InP wants to sell its resources.
	
	\textbf{RA-Request()}. Users authenticate the InP by checking its public key when they receive the transaction and confirm the available resources and the cost claimed by the InP (check($ \mathrm{WA_{ InP}} $, resources, cost)). If this information is correct, users will send their IDs, wallet addresses, quality of service (QoS) requirements, and SFC information in a transaction ($ ID_{\mathrm{ SFC}} $, $ \mathrm{WA_{ SFC}} $, requirement, information). Each user's ID is added to the list of SFCs (add(SFCs, $ ID_{\mathrm{ SFC}} $)).
	
	\textbf{ARA() \& HuRA()}. Upon receiving users' transactions, the InP invokes the corresponding functions of one of the two proposed algorithms in Section \ref{ARA HuRA} to address NFV-RA problem \eqref{nfv-problem} i.e., ARA() or HuRA() algorithms. It then authenticates users by checking their public keys (verify($ ID_{\mathrm{ SFC}} $, $ \mathrm{WA_{ SFC}} $)). After successful authentication, InP executes ARA() or HuRA() algorithm (run(ARA algorithm) or run(HuRA algorithm)). The resource allocation  results and the total cost of resources are sent to each user (send($ ID_{\mathrm{ SFC}} $,RA results) and send($ ID_{\mathrm{ SFC}} $,cost)).
	
	\textbf{Payment()}. Each user confirms InP's transaction by checking whether the QoS requirement is met (check(requirement satisfying)). It then checks the total cost declared by the InP (check(cost)). If the resource allocation results and total cost are correctly stated by InP, the user pays the cost to the InP (send($ \mathrm{WA_{ InP}} $, cost)).
	
	  Briefly, in NFVChain, each SFC requesting user sends a transaction to InP, including its SFC information such as number and type of VNFs, required CPU cycles for each VNF, and required bandwidth between VNFs. Then,  InP performs ARA or HuRa algorithms to allocate resources to SFC requesting users. The resource allocation results are sent to each user. These transactions are aggregated to one block by miners. Each miner who generates a block should perform the mining process. The winner miner's block is added to BC. After adding the block to BC, all transaction such as transactions including resource allocation results are performed.

	To add a block to BC, the  transactions generated by InP and users  should be verified by the miner users. To do so, the miner users check whether all transactions meet all the conditions stated in the smart contract or not. If all conditions are met, the transaction is verified. To verify each transaction, miner users should check (1) the QoS requirement of users are satisfied, (2) the total cost announced by InP should be obtained according to the unit price of resources, (3)  resource allocation algorithms are implemented correctly and their outputs are correct, and (4) the cost paid by the user should be the same as the cost declared by the InP. After transaction verification, miner users place a number of verified transactions in a new block and start the mining process. Once the block is mined,  the new block is broadcast to all other miner users for verification, and the miner users add it to the BC after block verification. Each block contains a transaction in which the miner user has specified its preferred reward. Verification of the blocks is done in terms of %some metrics; that is 
	(1) miner user's requested reward not exceeding the specified value in the BC protocol, (2) all transactions within the block have been previously verified, and (3) the found nonce through the mining process is correct and is less than the target value. When a block is added to the BC,  by referring to the BC, users can inform what resources are available from the data center.% and what resources have been used. 
	\vspace{-1.35 em} 
	\subsection{Blockchain Network Notations}
	The mining process is a high computational task, since in this paper, the mining is done by resource-limited mobile devices, we assume the miner users offload the mining task to a group of users so that each of them performs a part of this task.
	In our system model, there is a set of $ \mathcal{U_M} $ miner users.    Assuming the size of mining task of miner user $ i\in \mathcal{U_M} $ is equal to  $ D_i $ bits and each bit  requires $ C_i $ CPU cycles, miner user $ i $ offloads this task to a set of $ \mathcal{K}_i $ users  where $ \mathcal{K}_i\subseteq \mathcal{U} $ \footnote{ The participating users in mining process may be selected from all users in system i.e., $ \mathcal{U} $.}.  To represent  the ratio of mining task $ i $ allocated to user $ k\in \mathcal{K}_i $, we define a continuous weighting variable $ f_{i,k} $, where $ \sum\limits_{k \in \mathcal{K}_i}f_{i,k}=1,~\forall i \in \mathcal{U_M} $. In this case, miner user $ i $ should pay to the other users for consuming their processing capacity. Therefore, for each CPU cycle, the miner user $ i $ pays  $ \mathrm{cost}_{i,k} $ to user $ k $ for participating in the mining process.
	Miner user $ i $ should transmit a portion of the task that  user $ k $ must process to it.  For transmission, user $ i $ consumes energy which is calculated as the product of power consumption and  transmission time. Assume each miner user communicates with other users on an orthogonal channel.  The transmission time for sending a portion of mining task to user $ k $ is obtained by $ T_{i,k}^{tr}= {f_{i,k}D_i}/{ R_{i,k}}$ where $ R_{i,k} $ is the data rate between  miner user $ i $ and user $ k $ \cite{mining-lambda}, \cite{mining-reward}. Assuming a constant transmit power for all miner users, $ R_{i,k} $ is calculated by $ R_{i,k}=\log_2\left(1+\dfrac{p_{i,k}h_{i,k}}{\sigma_k}\right) $ where $ p_{i,k} $ is the transmit power of miner user $ i $ to user $ k $, $ h_{i,k} $ is the path-gain from miner user $ i $ toward user $ k $, and $ \sigma_k $ is the noise power at user $ k $. Accordingly, the energy consumption for transmission from miner user $ i $ to user $ k $ is given by $ p_{i,k}(f_{i,k}D_i / R_{i,k}) $. 
	The taking time  of user $ k $ to process the corresponding portion of mining task $ i $ is given by $ T_{i,k}^{proc}=f_{i,k}D_iC_i/F_k^{\mathrm{max}} $ where $ F_k^{\mathrm{max}} $ is the maximum CPU cycles per second of user $ k $'s device. Hence, the energy consumption of user $ k $ for mining is calculated by $ \widetilde{p}_k (f_{i,k}D_iC_i/F_k^{\mathrm{max}}) $, where $ \widetilde{p}_k $ is power consumption of user $ k $'s device for processing at each second. %Also, miner $ i $ may perform a portion of  mining task itself, so the energy consumed by  miner $ i $ to perform the corresponding portion must be taken into account in the energy consumed by the mining process. The energy consumption of miner $ i $ to perform the mining process is obtained by $ \widetilde{p}_i \dfrac{f_{i,i}}{F_i^{\mathrm{max}}} $.
	Accordingly, the total energy consumption during mining process is calculated by
	 \begin{equation}\label{mining-energy}
	\begin{aligned}
	E_{\mathrm{mine}}\!=\!\!\!\sum\limits_{i\in\mathcal{U_M}}\sum\limits_{k \in \mathcal{K}_i}\!\!\left[p_{i,k}\left(\dfrac{f_{i,k} D_i}{R_{i,k}}\right)+\widetilde{p}_k \left(\dfrac{f_{i,k}D_iC_i}{F_k^{\mathrm{max}}} \right)\right].
	\end{aligned}
	\end{equation} 
	\vspace{-2.25 em}
	%\section{Problem Statement }\label{problem statement}
	%Then, since resource allocation management is done by smart contracts with BC technology and the foundation of the BC is the mining process, the optimization problem for offloading the mining task on a group of users is then defined.
	\section{Problem Formulation for NFV Resource Allocation}\label{problem statement}
	In this section,  the optimization problem for allocating data center resources to  SFC requesting users is stated. To do so, the multi-objective optimization problem is formally defined to minimize data center energy consumption and minimize the cost of utilized  resources. In this problem, there are a number of constraints that must be satisfied. These constraints include constraints for VNF placement, routing, and the users' QoS, which are respectively explained in what follows.
	\subsubsection{VNF Placement Constraints}
	For embedding the SFCs at  data centers, only one server should be allocated to each VNF $ j\in\mathcal{S}_i $. Therefore, in the resource allocation problem, we have
	\begin{equation}\label{node-allocation}
	\begin{aligned}
	\mathrm{C1:}\sum\limits_{n\in\mathcal{N}} x_n^{i,j}=1,~~~~~\forall i\in\mathcal{U_S},~~~\forall j\in\mathcal{S}_i.
	\end{aligned}
	\end{equation}
	
	We assume that every VNF of each SFC should be mapped to a different server. So,  we have 
	\begin{equation}\label{node-allocation-different server}
	\begin{aligned}
	\mathrm{C2:}\sum\limits_{j\in\mathcal{S}_i} x_n^{i,j}\leq 1,~~~~~\forall i\in\mathcal{U_S},~~~\forall n\in\mathcal{N}.
	\end{aligned}
	\end{equation} 
	
	As aforementioned, each server at data center has limited processing capacity. The processing capacity limitation of servers which implement VNFs is represented by
	\begin{equation}\label{node-capacity}
	\begin{aligned}
	\mathrm{C3:}~~\sum\limits_{i\in\mathcal{U_S}} \sum\limits_{j\in\mathcal{S}_i}x_n^{i,j} C_{i,j}\leq\beta_n C_{n}^{\mathrm{max}},~~~~~\forall n\in\mathcal{N}.
	\end{aligned}
	\end{equation}
	
	The following constraint ensures that server $ n $ is active only when it hosts at least one VNF
	\begin{equation}\label{node-on}
	\begin{aligned}
	\mathrm{C4:}~~x_n^{i,j}\leq\beta_n ,~~~\forall n\in\mathcal{N},~~\forall i \in\mathcal{U_S},~~\forall j \in \mathcal{S}_i.
	\end{aligned}
	\end{equation}

	\subsubsection{Routing Constraints}
	Let $ \mathcal{L}_n^{\mathrm{out}} $ and $ \mathcal{L}_n^{\mathrm{in}} $ denote the outgoing links from server $ n $ and incoming links to server $ n $, respectively.  The following constraint enforces flow conservation, i.e., the incoming traffic in  servers that do not host VNFs should be equal to the outgoing ones. More precisely, this constraint ensures that the sum of the outgoing links bandwidth of server $ n $ that is running $ j $th VNF of SFC $\mathcal{S}_ i $ has to be the same as required bandwidth for transmitting traffic from  $ j $th to $ j + 1 $th VNF.  Also, this constraint makes sure that sum of the incoming links bandwidth of server $ n $ that runs $ j+1 $th VNF of SFC $\mathcal{S}_ i $ has to be the same as required bandwidth for traffic transmission from  $ j $th to $ j + 1 $th VNF. So, the flow conservation constraint can be considered in the resource allocation problem as
	\begin{equation}\label{flow conservation}
	\begin{aligned}
	&\mathrm{C5:}~~\sum\limits_{l\in\mathcal{L}^{\mathrm{out}}_{n}}~y_{l_j^{j+1}}^i-\sum\limits_{l\in\mathcal{L}^{\mathrm{in}}_{n}}~y_{l_j^{j+1}}^i=B^{j,j+1}_i(x_n^{i,j}-x_n^{i,j+1}) ,\\&\forall i \in \mathcal{U_S},~~~\forall j\in\mathcal{S}_i\cup \{0\},~~~\forall n\in\mathcal{N}.
	\end{aligned}
	\end{equation}
	It should be noted that  constraint $\mathrm{C5}$ ensures that sum of the outgoing links bandwidth of access switch (source node)  to the server on which AMF is implemented  has to be the same as required bandwidth between  access switch and AMF. Likewise, the incoming links bandwidth to  the transport switch (destination node) from UPF host server is the same as required bandwidth between UPF and transport switch.

	The maximum bandwidth of links carry out the traffic is indicated by
	\begin{equation}\label{link-cp}
	\begin{aligned}
	\mathrm{C6:}~~\sum\limits_{i\in\mathcal{U_S}}\sum\limits_{j\in\mathcal{S}_i\cup\{0\}}y_{l_j^{j+1}}^i\leq B_{l}^{\mathrm{max}},~~~\forall l\in\mathcal{L},
	\end{aligned}
	\end{equation}

	\subsubsection{SFC Requesting Users' QoS}
	The QoS requirement for user $ i $'s SFC is defined as maximum  end-to-end tolerable delay, $ T_i^{\mathrm{th}} $. Accordingly, we have 
	  \begin{equation}\label{delay-tolerable}
	\begin{aligned}
	&\mathrm{C7:}\sum\limits_{n \in \mathcal{N}}\sum\limits_{j \in \mathcal{S}_i}x_n^{i,j}(\dfrac{C_{i,j}}{C_n^{\mathrm{max}}}) +\sum\limits_{l \in \mathcal{L}} \sum\limits_{j \in \mathcal{S}_i\cup\{0\}} \dfrac{ y_{l_j^{j+1}}^i}{B_l^{\mathrm{max}}}\leq T_i^{\mathrm{th}},\\
	&\forall i\in\mathcal{U_S}.
	\end{aligned}
	\end{equation}    
	$ \mathrm{C7} $ means that the summation of processing delay on servers and transmission delay on physical links should not be  larger than a maximum tolerable delay.
	
	In  NFV-RA, minimizing data center energy consumption and minimizing the cost of utilized data center resources are of interest. The energy consumed ($ E $) in the data center is given by
	  \begin{equation}\label{energy-consumption}
	\begin{aligned}
	E=\sum\limits_{n \in \mathcal{N}} \beta_np_n^s+\sum\limits_{n \in \mathcal{N}}p_n\sum\limits_{i\in\mathcal{U_S}}\sum\limits_{j \in \mathcal{S}_i}(x_n^{i,j}C_{i,j}/C_n^{\mathrm{max}}).
	\end{aligned}
	\end{equation}  
	The cost of resources for user $ i $'s SFC is expressed as the unit price of each  CPU cycle of server $ n $ denoted by $ \mathrm{cost}_{i,n} $ and the unit price for transmitting  each bit per second  over physical link $ l $ denoted by $ \mathrm{cost}_{i,l} $. So, the cost of resources  used by  all  SFC requesting users is obtained by
	
	  \begin{equation}\label{cost-per-user}
	\begin{aligned}
	&\mathrm{cost}=\sum\limits_{n \in \mathcal{N}} \sum\limits_{i\in\mathcal{U_S}}\sum\limits_{j \in \mathcal{S}_i}(\mathrm{cost}_{i,n}x_n^{i,j}C_{i,j})\\&+ \sum\limits_{l \in \mathcal{L}}  \sum\limits_{i\in\mathcal{U_S}} \sum\limits_{j \in \mathcal{S}_i\cup \{0\}}(\mathrm{cost}_{i,l}y^i_{l_j^{j+1}}).
	\end{aligned}
	\end{equation}

	Accordingly, the objective function of NFV-RA problem is expressed as
	 \begin{equation}\label{objective-function}
	\begin{aligned}
	&\displaystyle \min_{\substack{\boldsymbol{\beta},\boldsymbol{X},\boldsymbol{Y}}}
	&& \sum\limits_{n \in \mathcal{N}} \beta_np_n^s+\sum\limits_{n \in \mathcal{N}}p_n\sum\limits_{i\in\mathcal{U_S}}\sum\limits_{j \in \mathcal{S}_i}(x_n^{i,j}C_{i,j}/C_n^{\mathrm{max}}),\\
	&\displaystyle \min_{\substack{\boldsymbol{\beta},\boldsymbol{X},\boldsymbol{Y}}}
	&& \Big(\sum\limits_{n \in \mathcal{N}}\! \sum\limits_{i\in\mathcal{U_S}}\!\sum\limits_{j \in \mathcal{S}_i}\!\mathrm{cost}_{i,n}x_n^{i,j}C_{i,j}\!\\
	&
	&&+\!\! \sum\limits_{l \in \mathcal{L}} \! \sum\limits_{i\in\mathcal{U_S}} \!\sum\limits_{j \in \mathcal{S}_i\cup \{0\}}\!\mathrm{cost}_{i,l}y^i_{l_j^{j+1}}\Big).
	\end{aligned}
	\end{equation} 
	  The multi-objective function can be shown by a single objective function using the weighted sum method in which two objective functions are linearly combined \cite{multi-objective}. So, the objective function of NFV-RA problem in \eqref{objective-function} is illustrated by a single-objective function as    
	 \begin{equation}\label{single-objective-function}
	\begin{aligned}
	& F(\boldsymbol{\beta},\boldsymbol{X},\boldsymbol{Y})=\alpha\Big[\!\!\!\sum\limits_{n \in \mathcal{N}} \beta_np_n^s+\!\!\!\sum\limits_{n \in \mathcal{N}}p_n\sum\limits_{i\in\mathcal{U_S}}\sum\limits_{j \in \mathcal{S}_i}(x_n^{i,j}C_{i,j}/C_n^{\mathrm{max}})\Big]\\
	&+(1-\alpha)\Big[\sum\limits_{n \in \mathcal{N}}\ \sum\limits_{i\in\mathcal{U}}\sum\limits_{j \in \mathcal{S}_i}(\mathrm{cost}_{i,n}x_n^{i,j}C_{i,j})
	\\&+ \sum\limits_{l \in \mathcal{L}}  \sum\limits_{i\in\mathcal{U}} \sum\limits_{j \in \mathcal{S}_i\cup \{0\}}(\mathrm{cost}_{i,l}y^i_{l_j^{j+1}} )\Big],
	\end{aligned}
	\end{equation} 
	  where $ 0\leq \alpha \leq 1 $ is a given weighted factor which reflects the relative importance of energy consumption and utilized resources cost \cite{multi-objective} , $ \boldsymbol{\beta} $ is  a vector of binary values which indicates the active and inactive modes of servers, $ \boldsymbol{X} $ and $ \boldsymbol{Y} $ are the server and link bandwidth allocation matrices. 
	Therefore, the NFV-RA problem is formally stated as 
	\begin{equation}\label{nfv-problem}
	\begin{aligned}
	&\displaystyle \min_{\substack{\boldsymbol{\beta},\boldsymbol{X},\boldsymbol{Y}}}
	&&F(\boldsymbol{\beta},\boldsymbol{X},\boldsymbol{Y})\\
	&\text{s.t.}
	&& ~\mathrm{C1}, ~\mathrm{C2}, ~\mathrm{C3}, ~\mathrm{C4}, ~\mathrm{C5}, ~\mathrm{C6}, ~\mathrm{C7},\\
	&
	&&\mathrm{C8:} ~~\beta_n \in \{0,1\},~\forall n\in\mathcal{N},\\
	&
	&&\mathrm{C9:}~~ x_n^{i,j} \in \{0,1\},~\forall n\in\mathcal{N},~\forall i \in\mathcal{U_S},~\forall j \in \mathcal{S}_i,\\
	&
	&&\mathrm{C10:}~~ y^i_{l_j^{{j+1}}} \geq 0,~\forall l\in\mathcal{L},~\forall i \in\mathcal{U_S},~\forall j \in \mathcal{S}_i\cup \{0\},
	\end{aligned}
	\end{equation}
	where $ \mathrm{C8} $ and  $ \mathrm{C9} $ represents the binary nature of servers' modes and server allocation variables, respectively. $ \mathrm{C10}$ implies that the allocated links bandwidth to users' SFC should be a non-negative value. 
	Problem \eqref{nfv-problem} is an MILP problem and generally   NP-hard. In Section \ref{ARA HuRA}, we propose two sub-optimal algorithms named ARA and HuRA to address problem \eqref{nfv-problem}.
	%*************************************************************
	\vspace{-0.8 em}
	\section{Problem Formulation for Mining Process}\label{mining}

	  To allocate NFV resources to SFC requesting users in a distributed manner employing BC technology, miner users should perform the mining process. The mining process consumes a significant amount of energy, so minimizing the energy consumption in the mining process is an important challenge.  Besides, since in wireless networks, the resource-limited and battery-powered mobile devices act as miners,  they cannot perform the mining process on their own, so in this paper, we assume that miner $ i $ offloads the processing required for the mining process to a group of users and pays to them for their processing capacity consumption. Also, any miner who mines the block faster than the others receives a reward from BC network. Note that the communication between miners and participating users is provided through the device to device communication.   
	
	The reward that the winning miner receives is made up of two components: (1) a constant value ($ R_{\mathrm{const}} $), (2) a variable value that depends on the number of transactions in the block ( $ N_{\mathrm{Trans}}R_{\mathrm{Trans}} $, where $ N_{\mathrm{Trans}} $ is the number of transactions in the mined block and $ R_{\mathrm{Trans}} $ is reward of each transaction) \cite{mine-jiot-2019},  \cite{mine-jiot-2019-2}. 
	
	The successfully of a mined block depends on two steps. In the first step, the winning miner must finish the mining process faster than the other miners. The probability of successfully at this step depends on the relative computing power of the miner. This probability is obtained from the ratio of the miner user  $ i $'s demand to the total demand of all miner users as $ D_iC_i/\sum\limits_{j \in \mathcal{U_M}}D_jC_j $ \cite{mine-jiot-2019}, \cite{mining-reward}. In the second step, the winning block should be propagated faster than the other blocks in the network. A block may mine quickly, but because of its large size, it will be discarded due to the long propagation latency, which is called orphaning. The probability of a block being orphaned depends on the number of transactions in the block. It is also assumed that mined block production follows the Poisson distribution. Accordingly, the probability of a block orphaning is given by $ p_{\mathrm{orphan}}=1-e^{-\lambda z N_{\mathrm{Trans}}} $, where  $ \lambda=\frac{1}{600}  $ is the mean value of Poisson distribution\footnote{ Since in BC networks, at a rate of every $ 600 $ seconds, a new block is generated, the  difficulty of generating  a new block is dynamically adjusted 	so that it takes $ 600 $ seconds.  Accordingly, the	mining Poisson process has a fixed parameter	for the whole  miner users which equals to $ \lambda=\frac{1}{600}  $ \cite{mining-lambda}. }  and $ z $ is  a given network latency parameter. So the probability of winning a block is $ 1-p_{\mathrm{orphan}}=e^{-\lambda z N_{\mathrm{Trans}}} $ \cite{mine-jiot-2019}, \cite{mine-jiot-2019-2}. 
	The corresponding reward to the miner $ i $ is obtained by 
	
	\begin{equation}\label{reward}
	\begin{aligned}
	Rw_i=\dfrac{D_iC_i}{\sum\limits_{j \in \mathcal{U_M}}D_jC_j}(R_{\mathrm{const}}+N_{\mathrm{Trans}}R_{\mathrm{Trans}})e^{-\lambda z N_{\mathrm{Trans}}}.
	\end{aligned}
	\end{equation}
	In this paper, we aim at minimizing the energy consumption of mining process and minimizing the cost minus reward of miners. Accordingly, the objective function of MO problem is
	 \begin{equation}\label{objective-mining}
	\begin{aligned}
	&\displaystyle \min_{\substack{\boldsymbol{f}}}
	&& \sum\limits_{i\in\mathcal{U_M}}\sum\limits_{k \in \mathcal{K}_i}\left[p_{i,k}\left(\dfrac{f_{i,k}D_i }{R_{i,k}}\right)+\widetilde{p}_k \left(\dfrac{f_{i,k}D_iC_i}{F_k^{\mathrm{max}}}\right) \right],\\
	&\displaystyle \min_{\substack{\boldsymbol{f}}}
	&& \sum\limits_{i \in \mathcal{U_M}}\sum\limits_{k \in \mathcal{K}_i} \mathrm{cost}_{i,k} f_{i,k}D_iC_i -\sum\limits_{i \in \mathcal{U_M}}Rw_i .
	\end{aligned}
	\end{equation} 
	Similar to the objective function of NFV-RA, \eqref{objective-mining} can be illustrated as a single objective function 
	 \begin{equation}\label{objective-mining-single}
	\begin{aligned}
	&\displaystyle \min_{\substack{\boldsymbol{f}}}G\left(\boldsymbol{f}\right)=\gamma\!
	\left(\sum\limits_{i\in\mathcal{U_M}}\sum\limits_{k \in \mathcal{K}_i}\left[p_{i,k}\!\left(\dfrac{f_{i,k}D_i }{R_{i,k}}\right)\!+\!\widetilde{p}_k\! \left(\dfrac{f_{i,k}D_iC_i}{F_k^{\mathrm{max}}}\right) \right]\right)\\&+\left(1-\gamma\right) \left(\sum\limits_{i \in \mathcal{U_M}}\sum\limits_{k \in \mathcal{K}_i} \mathrm{cost}_{i,k} f_{i,k}D_iC_i -\sum\limits_{i \in \mathcal{U_M}}Rw_i\right) ,
	\end{aligned}
	\end{equation} 
	where $ 0\leq \gamma \leq 1 $ is a weighted factor which reflects the relative importance of energy consumption and offloading cost \cite{multi-objective} and $ \boldsymbol{f} $ is the offloading matrix.
	Hence, to implement NFVO in a distributed manner, the MO problem for mining process is formally stated as	
	 \begin{subequations}\label{problem-mining}
		\begin{align}
		&\displaystyle \min_{\substack{\boldsymbol{f}}}
		&&G(\boldsymbol{f})\\
		&\text{s.t.}
		\label{sum of all offloading}
		&&\sum\limits_{k \in \mathcal{K}_i}f_{i,k}=1,~~~\forall i \in \mathcal{U_M},\\
		\label{capacity of each user}
		&
		&&\sum\limits_{i \in \mathcal{U_M}}f_{i,k}D_iC_i\leq F_k^{\mathrm{max}},~~~\forall k \in\mathcal{K}_i,~\forall \mathcal{K}_i \subseteq \mathcal{U},\\
		\label{orphaning-delay}
		&
		&&\displaystyle\max_{\substack{k \in \mathcal{K}_i}}\left(\dfrac{f_{i,k}D_i}{ R_{i,k}}+\dfrac{f_{i,k}D_iC_i}{F_k^{\mathrm{max}}}\right)\leq T_i^{\mathrm{ mine}},~\forall i \in \mathcal{U_M},\\
		\label{continoues offloading}
		&
		&&f_{i,k}\geq 0,~~~\forall i \in\mathcal{U_M},~~~\forall k \in\mathcal{K}_i,~~~\forall \mathcal{K}_i \subseteq \mathcal{U},
		\end{align}
	\end{subequations} 
	where \eqref{sum of all offloading} implies that the total portion of the mining task executed by all users collaborating to do mining task $ i $ should be equal to $ 1 $, \eqref{capacity of each user} represents that the total  CPU cycles of user $ k $ which is allocated to miner users to do mining task should be less than the maximum processing capacity of user $ k $'s device, \eqref{orphaning-delay} shows that the delay for performing a mining process should be less than a maximum delay. 
	
	  By solving the MO problem  \eqref{problem-mining}, each miner offloads its mining task onto a set of $ \mathcal{K}_i $ users so that each participating user performs a portion of the mining task. Since the decision variables $ f_{i,k}\geq 0,~\forall i \in\mathcal{U_M},~\forall k \in\mathcal{K}_i $ in  \eqref{problem-mining} is a continuous variable and the objective function and constraints of problem  \eqref{problem-mining} are linear functions with respect  to it, problem  \eqref{problem-mining} is an LP problem. Therefore, the optimal solution of problem  \eqref{problem-mining} can be easily obtained by off-the-shelf optimization software packages such as CVX toolbox  \cite{cvx}.  	
	%************************************************************
	\section{Our Proposed ARA and HuRA algorithms}\label{ARA HuRA}
	To solve NFV-RA problem \eqref{nfv-problem}, we obtain  optimal solution for a small-scale network using SCIP optimization  toolbox\footnote{[Online available at] \url{https://scip.zib.de/}}. Furthermore, to deal with NP-hardness of the MILP problem \eqref{nfv-problem}, we relax the binary variables and add a penalty function to the objective function of problem \eqref{nfv-problem} \cite{tvt-2019}, \cite{binary-relax}. This makes the objective function of problem \eqref{nfv-problem} non-convex, so, we approximate it by Majorization-Minimization approximation method \cite{MM-approximation} and obtain a near-optimal solution. This proposed algorithm is called ARA which has high computational complexity and is suitable for bench-marking. Therefore, to reduce the computational complexity of ARA, we propose a heuristic algorithm based on Hungarian algorithm named HuRA. 
	\subsection{Our Proposed ARA algorithm}
	Because of   binary nature of server allocation variables (i.e., $ x_n^{i,j} $) and active or inactive modes of servers (i.e., $ \beta_n $), NFV-RA problem \eqref{nfv-problem} is NP-hard. So, to overcome this difficulty, similar to \cite{tvt-2019} and \cite{binary-relax}, we replace the binary variables  constraint $ \mathrm{C8} $ and $ \mathrm{C9} $ in \eqref{nfv-problem}, respectively by  following equivalent constraints 
	\begin{equation}\label{relaxation-states}
	\begin{aligned}
	&\mathrm{C8.1:}~~\sum\limits_{n\in\mathcal{N}}(\beta_n-{\beta_n}^2)\leq 0,\\
	&\mathrm{C8.2:}~~0\leq \beta_n\leq 1,~~~\forall n\in\mathcal{N},
	\end{aligned}
	\end{equation}
	and
	
	\begin{equation}\label{relaxation}
	\begin{aligned}
	&\mathrm{C9.1:}~~\sum\limits_{n\in\mathcal{N}}\sum\limits_{i\in\mathcal{U_S}}\sum\limits_{j\in\mathcal{S}_i}(x_n^{i,j}-{x_n^{i,j}}^2)\leq 0,
	\\&\mathrm{C9.2:}~~0\leq x_n^{i,j}\leq 1, \forall n\in\mathcal{N}, \forall i \in\mathcal{U_S},\forall j\in\mathcal{S}_i.
	\end{aligned}
	\end{equation}
	By substituting the binary constraints $ \mathrm{C8} $  and $ \mathrm{C9} $  in \eqref{nfv-problem} with constraints $\mathrm{C8.1}$ and $\mathrm{C8.2}$ in \eqref{relaxation-states} and $\mathrm{C9.1}$ and $\mathrm{C9.2}$ in \eqref{relaxation},  problem \eqref{nfv-problem} is  transformed into a non-convex problem (due to the non-convexity of constraints $\mathrm{C8.1}$ and $\mathrm{C9.1}$). The following theorem is for handling constraints $\mathrm{C8.1}$ and $\mathrm{C9.1}$. 
	\begin{thm1}\label{theorem-1}
		For sufficiently large values for $ \lambda_1\gg 1 $ and $ \lambda_2\gg 1 $, problem \eqref{nfv-problem} is equivalent to the following problem.
		\begin{equation}\label{nfv-re-augment}
		\begin{aligned}
		&\displaystyle \min_{\substack{\boldsymbol{\beta},\boldsymbol{X},\boldsymbol{Y}}}
		&&F(\boldsymbol{\beta},\boldsymbol{X},\boldsymbol{Y})+ \lambda_1 \sum\limits_{n\in\mathcal{N}}(\beta_n-\beta_n^2)+\\&
		&&\lambda_2 \sum\limits_{n\in\mathcal{N}}\sum\limits_{i\in\mathcal{U_S}}\sum\limits_{j\in\mathcal{S}_i}(x_n^{i,j}-{x_n^{i,j}}^2)\\
		&\text{s.t.}
		&&\mathrm{C1},\mathrm{C2},\mathrm{C3},\mathrm{C4},  \mathrm{C5},\mathrm{C6},\mathrm{C7},\mathrm{C8.2},\mathrm{C9.2},\mathrm{C10},
		\end{aligned}
		\end{equation}
		where $ \lambda_1 $ and $ \lambda_2 $ act as penalty factors to penalize the objective function for any $ \beta_n $ and $ {x_n^{i,j}} $ that is not equal to $ 0 $ or $ 1 $. 
	\end{thm1}
	
	\begin{proof}
		The proof is given in the Appendix B.
	\end{proof}
	Let $ f(\boldsymbol{\beta,X,Y})=F(\boldsymbol{\beta},\boldsymbol{X},\boldsymbol{Y})+\lambda_1\sum\limits_{n \in \mathcal{N}}\beta_n+\lambda_2 \sum\limits_{n\in\mathcal{N}}\sum\limits_{i\in\mathcal{U_S}} \sum\limits_{j\in\mathcal{S}_i} x_n^{i,j}$ and $ g(\boldsymbol{\beta,X})=\lambda_1\sum\limits_{n \in \mathcal{N}}{\beta_n}^2+\lambda_2 \sum\limits_{n\in\mathcal{N}}\sum\limits_{i\in\mathcal{U_S}} \sum\limits_{j\in\mathcal{S}_i} {x_n^{i,j}}^2 $, the objective function of problem \eqref{nfv-re-augment} can be written as  difference of two convex functions $f(\boldsymbol{\beta,X,Y}) $ and $ g(\boldsymbol{\beta,X}) $. So, problem \eqref{nfv-re-augment} is a D.C programming problem. Majorization-Minimization approximation is a well-known  method to convexify a D.C programming problem as a convex one,  One approach for Majorization-Minimization approximation  is  the first-order Taylor approximation method. Thus, to convexify  the objective function of \eqref{nfv-re-augment}, we approximate $ g(\boldsymbol{\beta,X})=\lambda_1\sum\limits_{n \in \mathcal{N}}{\beta_n}^2+\lambda_2 \sum\limits_{n\in\mathcal{N}}\sum\limits_{i\in\mathcal{U_S}} \sum\limits_{j\in\mathcal{S}_i} {x_n^{i,j}}^2$ by its first-order Taylor approximation as  
	\begin{equation*}
	\begin{aligned}
	&g(\boldsymbol{\beta,X})=g({\boldsymbol{\beta}^{t-1}},{\boldsymbol{X}^{t-1}})+\nabla_ {\boldsymbol{\beta}}g({\boldsymbol{\beta}^{t-1}},{\boldsymbol{X}})(\boldsymbol{\beta}-{\boldsymbol{\beta}^{t-1}})\\&+\nabla_ {\boldsymbol{X}}g({\boldsymbol{\beta}},{\boldsymbol{X}^{t-1}})(\boldsymbol{X}-{\boldsymbol{X}^{t-1}}),
	\end{aligned}
	\end{equation*}
	where $ {\boldsymbol{\beta}^{t-1}} $ and $ {\boldsymbol{X}^{t-1}} $ is the optimal solution of previous iteration. 
	By doing so, problem \eqref{nfv-re-augment} can be rewritten as
	\begin{equation}\label{nfv-re-taylor}
	\begin{aligned}
	&\displaystyle \min_{\substack{\boldsymbol{\beta,X,Y}}}
	&& F(\boldsymbol{\beta},\boldsymbol{X},\boldsymbol{Y})+\lambda_1 \sum\limits_{n\in\mathcal{N}}\beta_n+\lambda_2 \sum\limits_{n\in\mathcal{N}}\sum\limits_{i\in\mathcal{U_S}}\sum\limits_{j\in\mathcal{S}_i}x_n^{i,j}\\
	&
	&&-\lambda_1\sum\limits_{n\in\mathcal{N}}\big[2{{\beta}_n}{{{\beta}_n}}^{t-1}-({{{\beta}_n}^{t-1})^2}\big]
	\\
	&
	&&-\lambda_2\sum\limits_{n\in\mathcal{N}}\sum\limits_{i\in\mathcal{U_S}}\sum\limits_{j\in\mathcal{S}_i}\big[2{{x}^{i,j}_n}{{{x}^{i,j}_n}}^{t-1}-({{{x}^{i,j}_n}^{t-1})^2}\big]\\
	&\text{s.t.}
	&&\mathrm{C1},\mathrm{C2},\mathrm{C3 },\mathrm{C4},  \mathrm{C5},\mathrm{C6},\mathrm{C7},\mathrm{C8.2},\mathrm{C9.2},\mathrm{C10}.
	\end{aligned}
	\end{equation}
	
	Problem \eqref{nfv-re-taylor}  is a convex optimization problem and its optimal  solution can be obtained by interior-point  \cite{boyd} method or using CVX toolbox \cite{cvx}.  To obtain a local optimum of NFV-RA problem \eqref{nfv-problem}, we employ an iterative algorithm to tighten the obtained upper bound as summarized in Algorithm 1. At each iteration, the convex problem in \eqref{nfv-re-taylor} is solved efficiently by  interior-point method. By solving the convex upper bound problem in \eqref{nfv-re-taylor}, the proposed iterative scheme generates a sequence of feasible solutions $ \boldsymbol{\beta}^{t} $, $ \boldsymbol{X}^{t} $, and $ \boldsymbol{Y}^{t} $ successively. The proposed suboptimal iterative algorithm converges to a locally optimal solution of problem \eqref{nfv-problem} in polynomial time.
	\begin{algorithm}\label{nfv-alg}
		\BlankLine

		\SetKwFunction{Range}{range}%%
		\SetKw{KwTo}{in}\SetKwFor{For}{for}{\string:}{}%
		\SetKwIF{If}{ElseIf}{Else}{if}{:}{elif}{else:}{}%
		
		\SetAlgoNoEnd
		
		\SetAlgoNoLine%
		\SetKwInOut{Input}{Input}
		\SetKwInOut{Output}{Output}
		Initialize  maximum number of iterations $ t^{\mathrm{max}} $,  $ \lambda_1, \lambda_2\gg 1 $,  iteration index $ t=1 $, and a feasible initial point  $ \boldsymbol{\beta}^1 $ and $ \boldsymbol{X}^1 $.\\
		\textbf{Repeat}\\
		\Indp
		Solve convex optimization problem \eqref{nfv-re-taylor} by interior-point method  \cite{boyd} and obtain {$ \boldsymbol{\beta}^* $, $ \boldsymbol{X}^* $,  $ \boldsymbol{Y}^* $}.\\
		Set  $ \boldsymbol{\beta}^t= \boldsymbol{\beta}^* $, $ \boldsymbol{X}^t= \boldsymbol{X}^* $, $ \boldsymbol{Y}^t= \boldsymbol{Y}^* $ and	 $ t\leftarrow t+1 $.\\
		\Indm
		\textbf{Until } convergence or $ t=t^{\mathrm{max}} $.
		\caption{Our proposed ARA algorithm  to solve NFV-RA  problem \eqref{nfv-problem}}	
	\end{algorithm}
	\begin{propos1}
		The optimal solution of problem \eqref{nfv-re-taylor} at each iteration $ t $ provides a tight upper bound and local optimal for problem \eqref{nfv-problem}.
	\end{propos1}
	\begin{proof}
		The proof is given in the Appendix C.
	\end{proof}
	\subsection{Our Proposed HuRA algorithm}
	In problem \eqref{nfv-re-taylor}, there are $ \mid\mathcal{U_S}\mid\mid\mathcal{S}_i\mid\mid\mathcal{N}\mid+\mid\mathcal{U_S}\mid\mid\mathcal{S}_i \mid\mid\mathcal{L}\mid+\mid\mathcal{N}\mid $ decision variables and $ \mid\mathcal{U_S}\mid \mid\mathcal{S}_i\mid\mid\mathcal{N}\mid+2\mid\mathcal{U_S}\mid \mid\mathcal{S}_i\mid\mid\mathcal{L}\mid+\mid\mathcal{U_S}\mid \mid\mathcal{S}_i\mid+\mid\mathcal{U_S}\mid \mid\mathcal{N}\mid+2\mid\mathcal{N}\mid+\mid\mathcal{L}\mid+\mid\mathcal{U_S}\mid $ linear constraints. So, the complexity of Algorithm 1 to solve problem \eqref{nfv-problem} is $ O((\mid\mathcal{U_S}\mid \mid\mathcal{S}_i\mid\mid\mathcal{N}\mid+\mid\mathcal{U_S}\mid \mid\mathcal{S}_i\mid\mid\mathcal{L}\mid+2\mid\mathcal{N}\mid)^2(\mid\mathcal{U_S}\mid \mid\mathcal{S}_i\mid+\mid\mathcal{U_S}\mid \mid\mathcal{N}\mid+\mid\mathcal{L}\mid+\mid\mathcal{U_S}\mid)) $ which is polynomial \cite{binary-relax}. In what follows, to reduce the computational complexity of our proposed ARA algorithm, we propose a heuristic algorithm based on Hungarian algorithm to address problem \eqref{nfv-problem} called as HuRA. To do so, we decompose  NFV-RA problem \eqref{nfv-problem} into two sub-problems, namely VNF placement and routing sub-problems. The VNF placement sub-problem is defined as 
	 \begin{equation}\label{vnf-placement}
	\begin{aligned}
	&\displaystyle \min_{\substack{\boldsymbol{\beta,X}}}~~~
	\alpha\Big[\sum\limits_{n \in \mathcal{N}} \beta_np_n^s+\sum\limits_{n \in \mathcal{N}}p_n\sum\limits_{i\in\mathcal{U_S}}\sum\limits_{j \in \mathcal{S}_i}x_n^{i,j}C_{i,j}/C_n^{\mathrm{max}}\Big]\\&+(1-\alpha)\Big[\sum\limits_{n \in \mathcal{N}}\ \sum\limits_{i\in\mathcal{U_S}}\sum\limits_{j \in \mathcal{S}_i}(\mathrm{cost}_{i,n}x_n^{i,j}C_{i,j}) \Big]\\
	&\text{s.t.}~\mathrm{C1},~\mathrm{C2},~\mathrm{C3},~\mathrm{C4},~\mathrm{C7},~\mathrm{C8},~\mathrm{C9},
	\end{aligned}
	\end{equation} 
	and the routing sub-problem is defined as follows.
	\begin{equation}\label{routing}
	\begin{aligned}
	&\displaystyle \min_{\substack{\boldsymbol{Y}}}~~~
	(1-\alpha)\Big[ \sum\limits_{l \in \mathcal{L}}  \sum\limits_{i\in\mathcal{U_S}} \sum\limits_{j \in \mathcal{S}_i\cup \{0\}}(\mathrm{cost}_{i,l}y^i_{l_j^{j+1}}) \Big]\\
	&\text{s.t.}~\mathrm{C5},~\mathrm{C6},~\mathrm{C7},~\mathrm{C10}.
	\end{aligned}
	\end{equation}
	The VNF placement problem \eqref{vnf-placement} is NP-hard due to the servers active or inactive modes and server allocation binary variables, so we propose a heuristic algorithm to solve it. Then, after placement the VNFs on the servers, the routing sub-problem  which is an LP problem is  optimally solved by the off-the-shelf optimization software packages.

	%\subsubsection{Solve the VNF Placement Sub-problem \eqref{vnf-placement}}
	Assuming having sufficient resources in data center to admit all SFCs, each VNF in each SFC must be implemented on one and only one server. To allocate servers to VNFs in each SFC, the Hungarian algorithm  which is an assignment algorithm can be employed \cite{hungarian}, \cite{hungarian-2}. To this end, first sort the SFCs in an  increasing order respect to their maximum tolerable delay ($ T_i^{\mathrm{th}} $). In each SFC $\mathcal{S}_ i $, for each VNF  $ j $, we add all servers that have sufficient remaining capacity to implement this VNF to candidate servers set  (i.e., $ \mathrm{ candidate}_{i,j}=[] $). Then, for each VNF $ j $ of SFC $\mathcal{S}_ i $, we calculate  the objective function of sub-problem \eqref{vnf-placement} for all the servers in  $ \mathrm{ candidate}_{i,j}$  and place it in  matrix $ A^{\mid\mathcal{N}\mid\times \mid\mathcal{N}\mid} $ \footnote{$ \mid\mathcal{N}\mid $ is the number of servers. Because each VNF in each SFC has to be assigned to a different server, the number of servers is greater than the number of VNFs in each chain.}.  $ A[j,n] $ represents the value of objective function if the server $ n $ is assigned to the $ j $th VNF of SFC $\mathcal{S}_ i $. If each server does not have the sufficient capacity  for a VNF, we set the value of the corresponding element in matrix $ A $ to a big  value that will not be selected by Hungarian algorithm.
	We give matrix $ A $ as an input to the Hungarian algorithm, the output of this algorithm is a matrix of $ 0 $ and $ 1 $ elements, indicating the assignment of the servers to the VNFs, so that the objective function of \eqref{vnf-placement} is minimized.
	%\subsubsection{Solve the Routing Sub-problem \eqref{routing}}
	After VNF placement, we solve the routing sub-problem to connect the servers run VNFs. By solving the routing sub-problem,  if  a feasible routing finds, the algorithm terminates. Otherwise, the VNF placement must be redone. In this case, for each VNF, we calculate the processing delay  for all the servers in the $ \mathrm{ candidate}_{i,j} $  and give the matrix as an input to the Hungarian algorithm. In this case, the output of the Hungarian algorithm will be a $ 0   $ and $ 1 $ matrix whose assignment is done in such a way that the processing delay for this SFC is minimized. Then the routing problem is resolved again. 
	
	After completion of VNF placement and routing for each SFC, we update the remaining capacity of the allocated servers as well as the bandwidth of allocated links. %To do so, the amount of servers capacity and links bandwidth used by this SFC deducts from the remaining servers capacity and links bandwidth, respectively. 
	Also, we update the value of objective function.
	It is worth noting that the computational complexity of  HuRA algorithm is $ O(\mid\mathcal{U_S}\mid\mid\mathcal{N}\mid^3+\mid\mathcal{U_S}\mid\mid\mathcal{S}_i\mid\mid\mathcal{N}\mid+\mid\mathcal{U_S}\mid\mid\mathcal{S}_i\mid\mid\mathcal{L}\mid) $, since the computational complexity of Hungarian algorithm is  $ O(N^3) $ \cite{hungarian-2} and Hungarian algorithm is implemented for each SFC, and the computational complexity to address LP problem \eqref{routing} is $ O(\mid\mathcal{U_S}\mid\mid\mathcal{S}_i\mid\mid\mathcal{N}\mid+\mid\mathcal{U_S}\mid\mid\mathcal{S}_i\mid\mid\mathcal{L}\mid)  $. The pseudo-code of our proposed HuRA algorithm is illustrated in	Algorithm 2 in Appendix D.
		
	%\section{Our Proposed Algorithm}\label{ARA HuRA}
	\section{Simulation Results}\label{simulation}
	In this section, we present the simulation results to evaluate the performance of our proposed  ARA and HuRA algorithms to address NFV-RA problem \eqref{nfv-problem}. To this end, we first compare the performance of the ARA and HuRA algorithms with the optimal solution to address problem \eqref{nfv-problem}. Then, we compare the ARA algorithm with the proposed algorithm  in \cite{rw-nfv-5} in terms of  the number of active servers, energy consumption, and average delay. Finally, we investigate the efficiency of the solution of MO problem \eqref{problem-mining} for the miner users. It is noteworthy that  all of the  curves  introduced in the following are obtained by averaging from  $ 100 $ independent snapshots. 
	\subsection{Comparison of Our Proposed ARA and HuRA algorithms with Optimal Solution}
	To evaluate our proposed ARA and HuRA algorithms,  we consider a directed graph in which servers are connected to each other with randomly established physical links. All  simulation parameters are given in Table \ref{simulation-parameter}. In what follows, we compare the performance of our proposed ARA and HuRA algorithms with the optimal solution of NFV-RA problem \eqref{nfv-problem}. %To obtain the optimal solution, we use SCIP optimization toolbox.
	\begin{table}[ht]
		\centering
		\caption{simulation parameters for NFV resource allocation } \label{simulation-parameter}
		
		\begin{tabular}{ |p{4 cm} | p{4 cm} |} 
			\hline\xrowht[()]{5 pt}
			\textbf{Parameter} & \textbf{Value}  \\  % inserts table
			%heading
			\hline 
			\hline \xrowht[()]{5 pt}
			$ \alpha $ & 0.5\\
			\hline \xrowht[()]{5 pt}						
			required bandwidth between $ j $ and $ j+1 $th VNF ($B_i^{j,j+1} $) & random selection of $[100,500]~\mathrm{bit/s}$ \\
			\hline \xrowht[()]{5 pt}
			required CPU cycles for VNF $ j $ & random selection of $[B_i^{j,j+1},5B_i^{j,j+1}]$\\
			 \hline\xrowht[()]{5 pt}
			unit price of each  CPU cycle  ($ \mathrm{cost}_{i,n} $)  & random selection of $[0.1,1]$ \\
			 \hline\xrowht[()]{5 pt}
			unit price for transmitted each bit per second  ($ \mathrm{cost}_{i,l} $) & random selection of $[0.1,1]$\\		
			 \hline\xrowht[()]{5 pt}
			server processing capacity ($C_n^{\mathrm{max}} $)	 &  random selection of $[1,10]~\mathrm{Mcpu~cycles/s}$ \\		
			 \hline\xrowht[()]{5 pt}
			link bandwidth ($B_l^{\mathrm{max}} $)		  &  random selection of $[100,500]~\mathrm{Mbps}$ \\
			 \hline\xrowht[()]{5 pt}
			power consumption 	of server $ n $ ($ {p}_n$)		      &    random selection $[1,5]~\mathrm{W}$ \\
			 \hline\xrowht[()]{5 pt}
			static power 	of server $ n $ ($ {p}_n^s$)		      &    random selection $[1,10]~\mathrm{W}$ \\
			 \hline\xrowht[()]{5 pt}
			VNFs number of each SFC ($ \mid \mathcal{S}_i \mid $) & random selection of $ (3,8) $\\
			\hline
		\end{tabular}
	\end{table}
	
	\begin{figure}
		\begin{center}
			
			\includegraphics[width=0.8\linewidth] {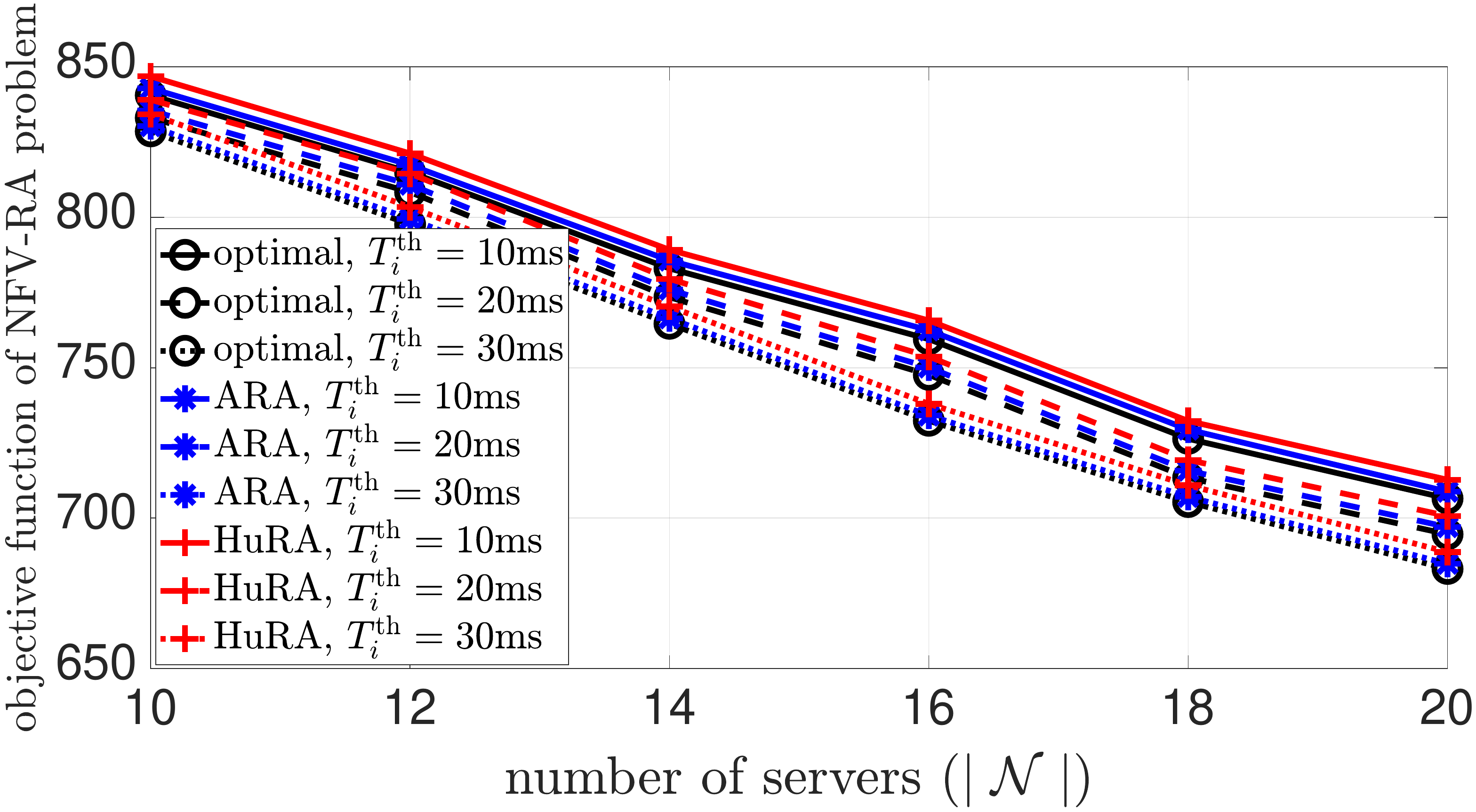}   
			\captionof{figure}{\small Objective function of NFV-RA problem \eqref{nfv-problem} vs.
				SFCs'  tolerable delay and number of servers}\label{delay_vs_servers} 
		\end{center}
	\end{figure} 
	
	\begin{figure}
		\begin{center}
			\includegraphics[width=0.8\linewidth]{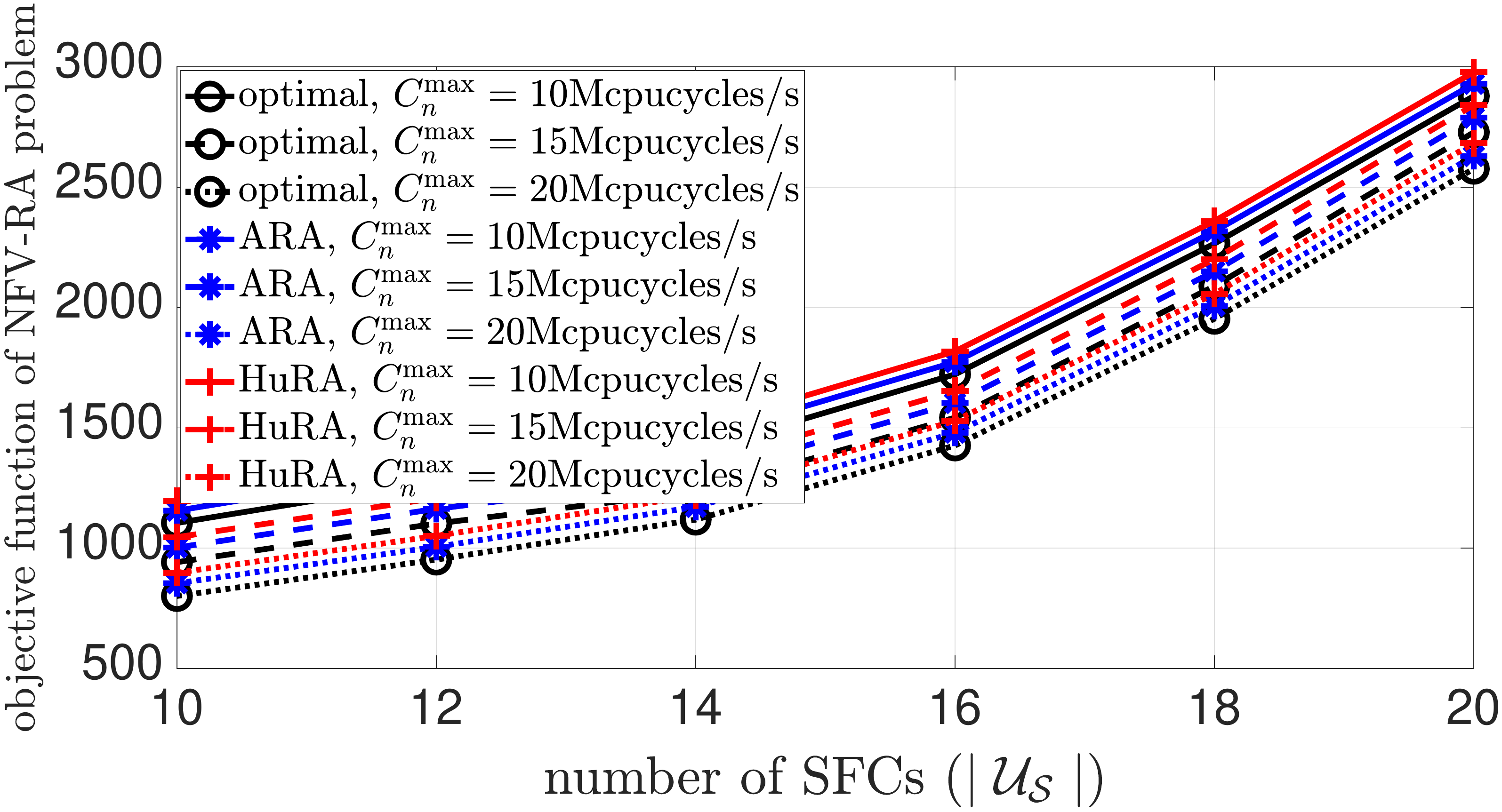}
			\captionof{figure}{ \small Objective function of NFV-RA problem \eqref{nfv-problem} vs.
				SFCs number and servers processing   capacity}\label{sfcs_vs_capacity}
			
		\end{center}
	\end{figure}
	
	Fig. \ref{delay_vs_servers} illustrates the performance of our proposed ARA and HuRA algorithms in comparison with the optimal solution of NFV-RA problem \eqref{nfv-problem} versus SFCs' maximum tolerable delay and the number of servers. For generating this figure, we set the number of SFCs to $ 5 $. From Fig. \ref{delay_vs_servers}, it can be seen that the objective function of NFV-RA problem \eqref{nfv-problem} decreases as the number of servers increases since there are more servers to choose for allocating to each VNF. Additionally, due to the processing capacity of the servers is randomly set, the probability of the existence of high-capacity servers increases, so we can run more VNFs on high-capacity servers. By doing so, the number of active servers is reduced which leads to a reduction in the objective function of problem \eqref{nfv-problem}.
	On the other hand, when the SFCs' maximum tolerable delay is reduced, the processing delay and transmission delay can be reduced. Therefore, servers with high processing capacity must be allocated to the VNFs of these SFCs, even if they require lower processing capacity. This means that due to the servers' processing capacity limitations, we cannot run VNFs of other SFCs that may require higher processing on these servers, and we will have to activate other servers. As a result, the number of active servers increases which results in an increase in the objective function of problem \eqref{nfv-problem}. Also, as can be seen in Fig. \ref{delay_vs_servers}, the ARA and HuRA algorithms achieve near-optimal performance with lower computational complexity.
	
	\begin{figure}
		\begin{center}
			
			\includegraphics[width=0.8\linewidth] {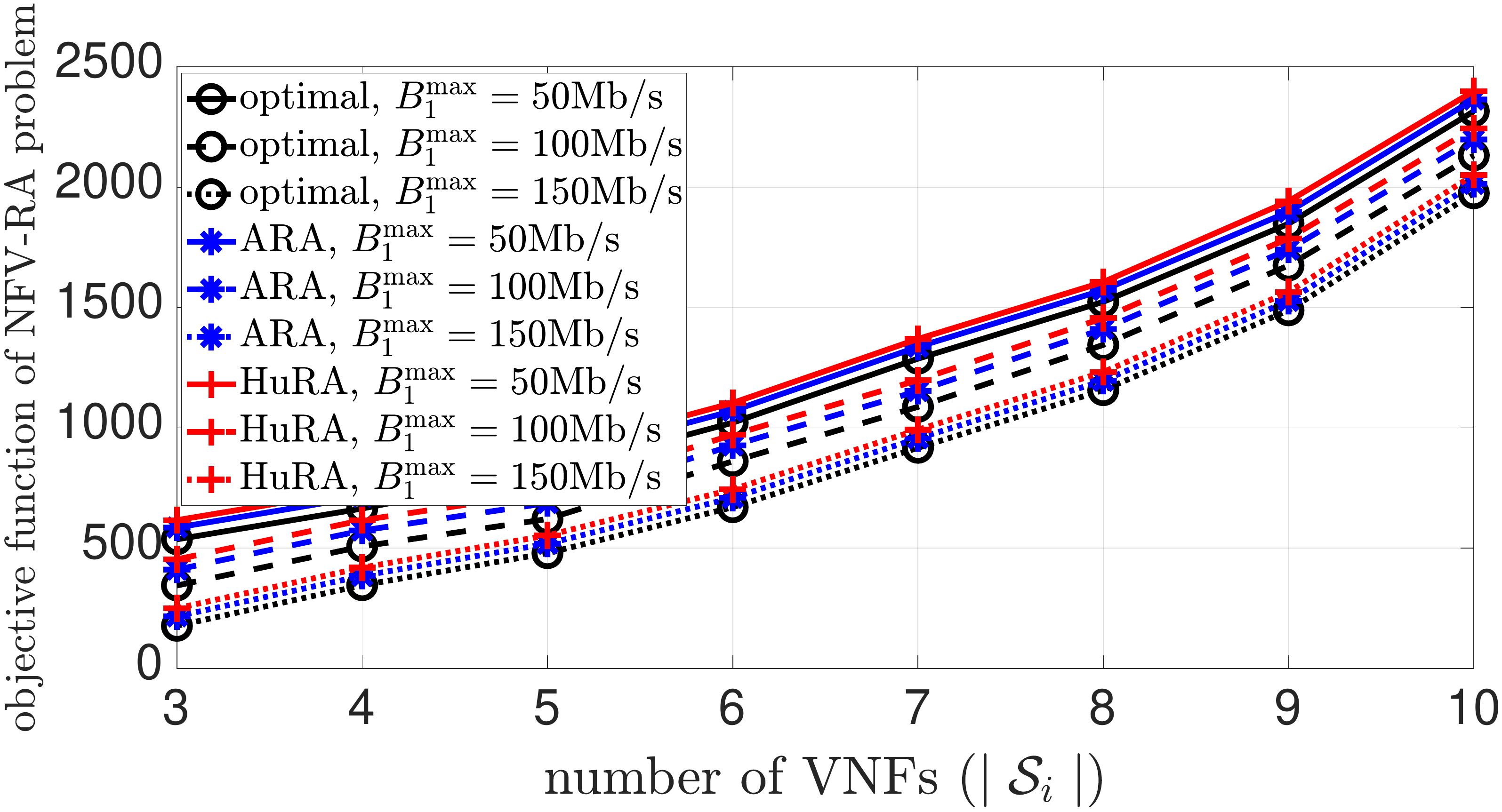}   
			\captionof{figure}{\small Objective function of NFV-RA problem \eqref{nfv-problem} vs.
				number of VNFs  and physical links bandwidth }\label{vnfs_vs_links}
			 
		\end{center}
	\vspace{-1.2 em}
	\end{figure}  
	\begin{figure}
		\begin{center}
			
			\includegraphics[width=0.8\linewidth]{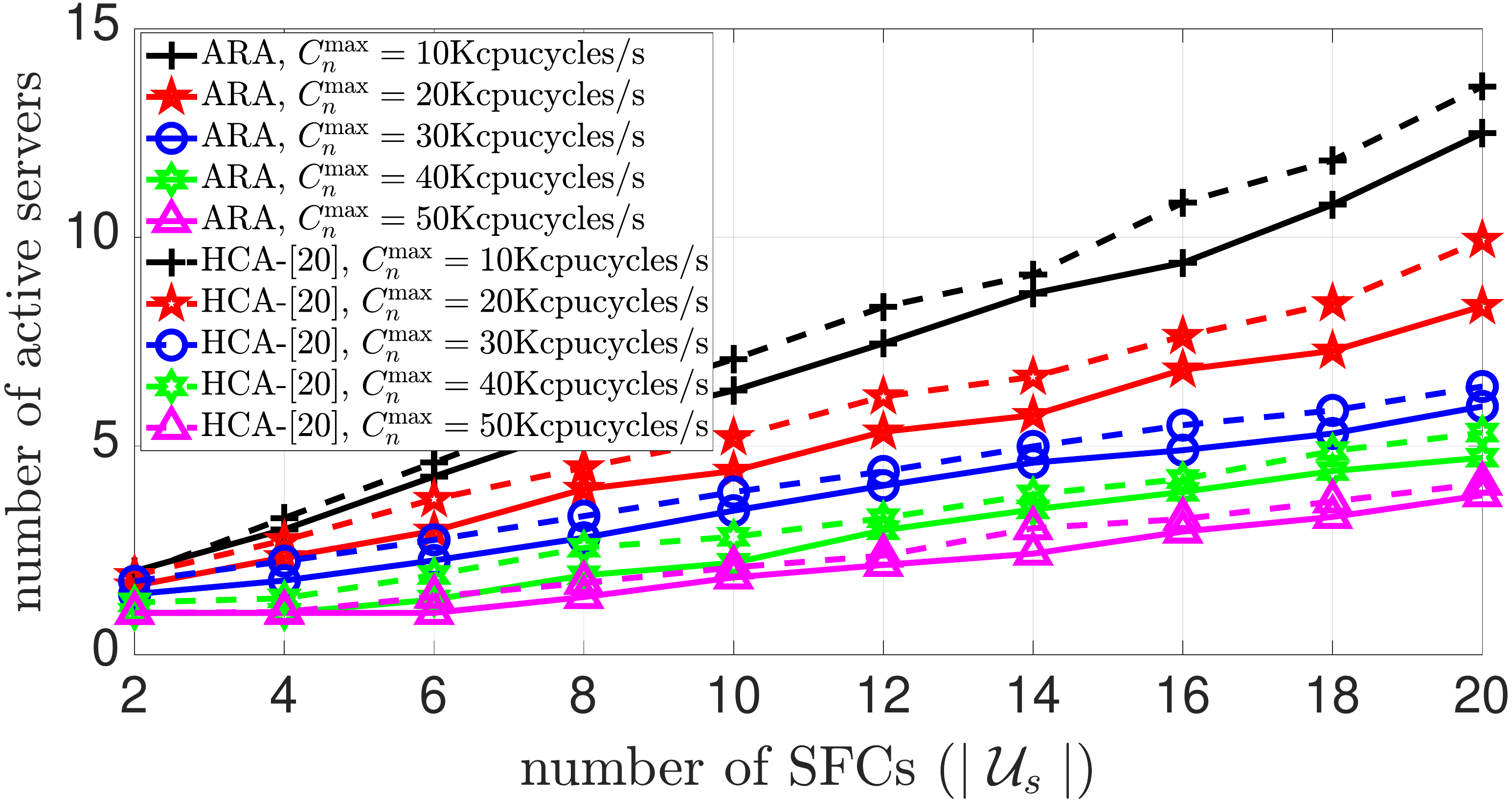}
			\captionof{figure}{ \small Number of active servers vs. number of SFCs	 and processing capacity of servers}\label{delay_vs_servers_comparison}
			 
		\end{center}
	\vspace{-1.2 em}
	\end{figure}
	\begin{figure}
	\begin{center}

		\includegraphics[width=0.8\linewidth]{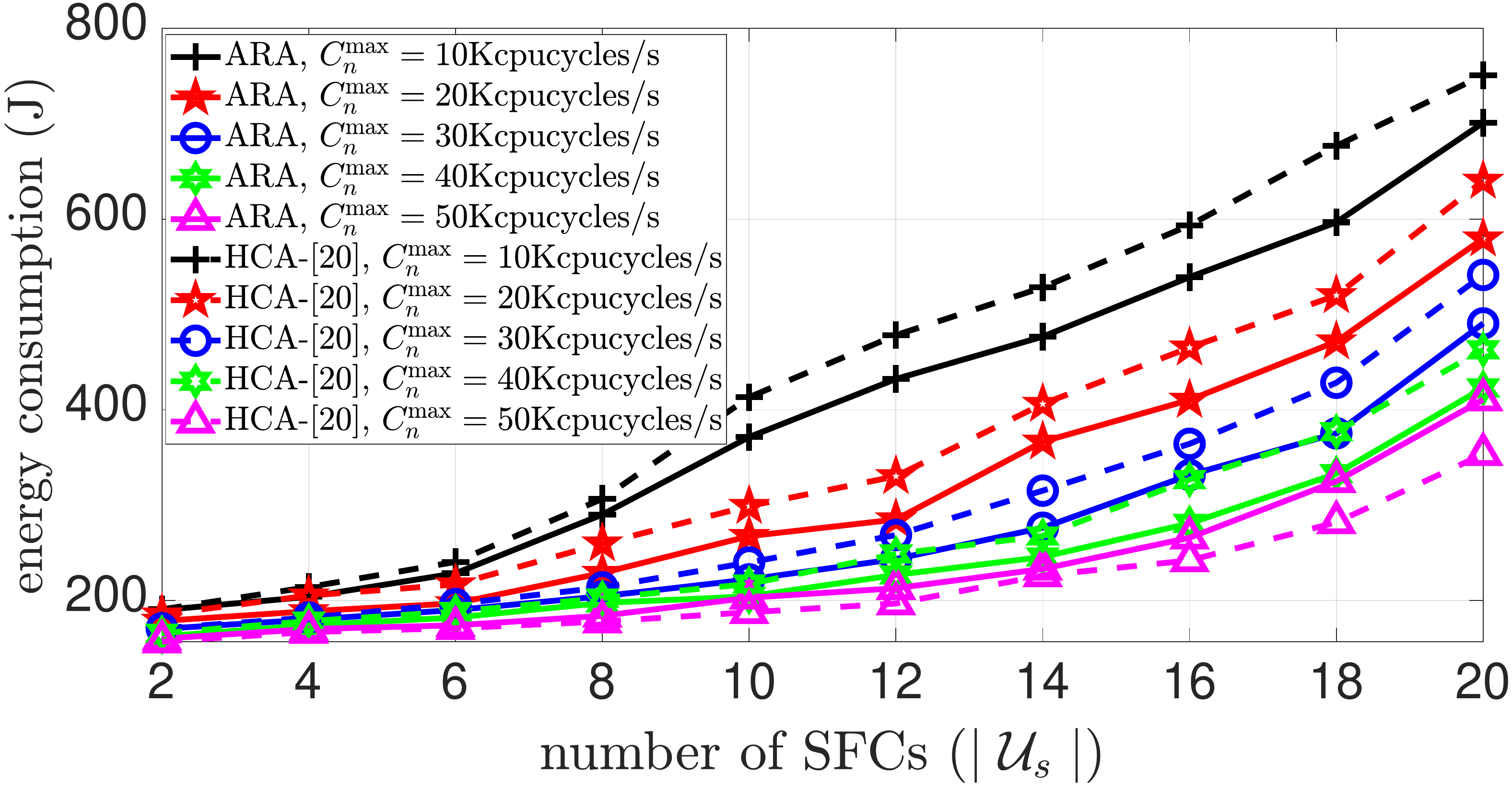}
		\captionof{figure}{ \small   Energy consumption vs. number of SFCs	 and processing capacity of servers }\label{energy_vs_sfcs_comparison}
		 
	\end{center}
\vspace{-1.2 em}
\end{figure}
	\begin{figure}
	\begin{center}
		
		\includegraphics[width=0.8\linewidth]{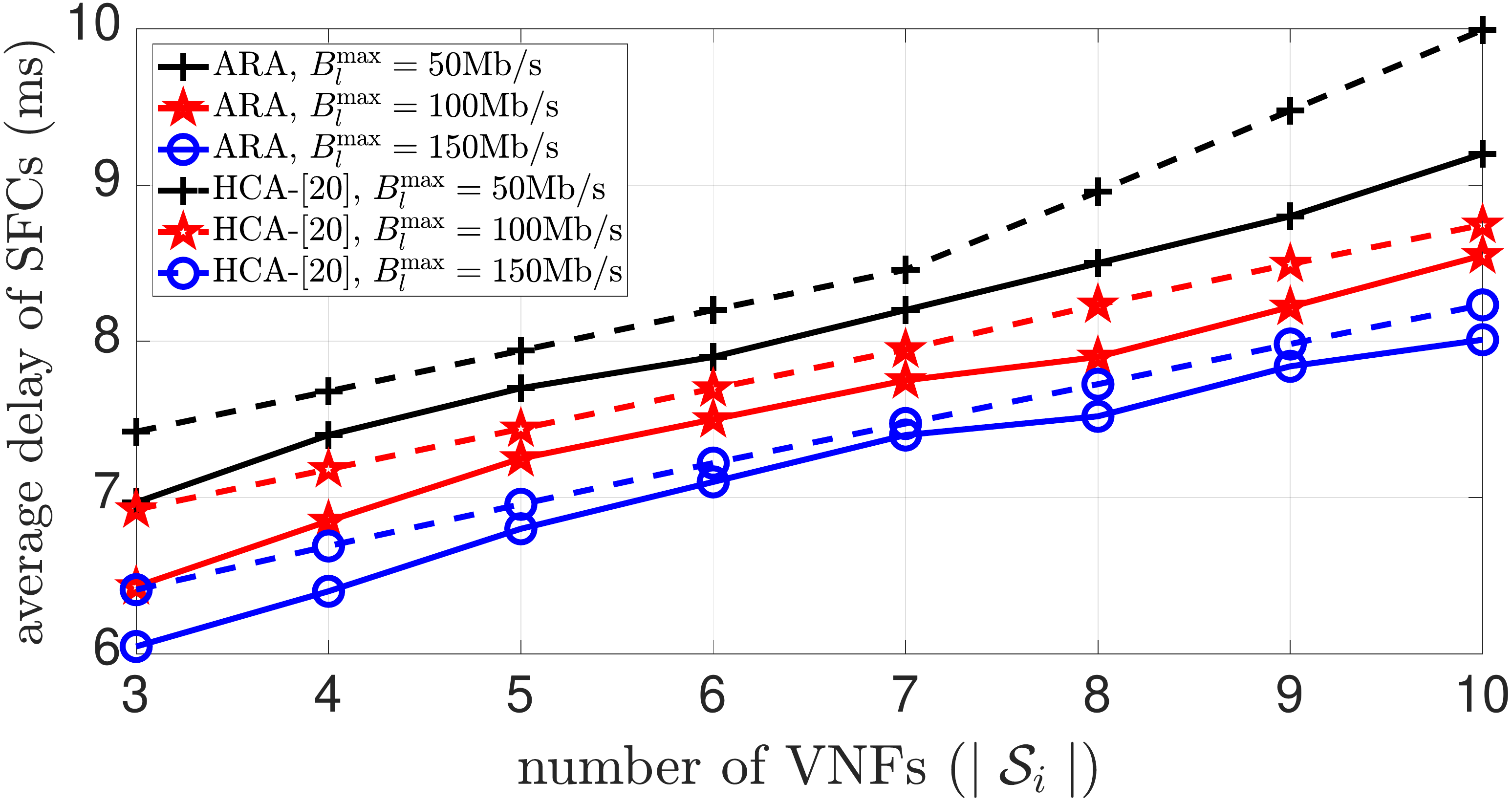}
		\captionof{figure}{ \small   Average delay vs. number of VNFs	 and bandwidth of physical links }\label{delay_vs_vnfs_comparison}
		 
	\end{center}
\vspace{-1.2 em}
\end{figure}

	The performance of our proposed ARA and HuRA algorithms in comparison with the optimal solution of problem \eqref{nfv-problem} versus the number of SFCs and processing capacity of servers is shown in Fig. \ref{sfcs_vs_capacity}. For generating this figure, we set the number of servers to $ 20 $ ($ \mid\mathcal{N}\mid=20 $) and SFCs' maximum tolerable delay to $ 20\mathrm{ ms} $ ($T_i^{\mathrm{th}}= 20\mathrm{ ms} $).
	It can be seen  that as the number of SFCs increases since more resources are needed for providing service to them and more consumed resources lead to increased energy consumption and cost of utilized resources, the objective function of  problem \eqref{nfv-problem}   increases. In addition, with increasing the processing capacity of servers, more VNFs can be implemented on each server, so the number of active servers is reduced. Also, as the servers' processing capacity increases, since the processing delay of the SFCs decreases, the energy consumption of the servers would be decreased. Thus, increasing the capacity of the servers results in decreasing the objective function of problem \eqref{nfv-problem}.

	Fig. \ref{vnfs_vs_links} shows the performance of our proposed  ARA and HuRA algorithms in comparison with the optimal solution of problem \eqref{nfv-problem} when the number of VNFs for each SFC as well as the bandwidth of physical links increases.  To generate this figure, we set the number of servers to $ 20 $ ($ \mid\mathcal{N}\mid=20 $), SFCs' maximum tolerable delay to $ 20\mathrm{ ms} $ ($T_i^{\mathrm{th}}= 20\mathrm{ ms} $), and the number of SFCs to $ 5 $ ($ \mid\mathcal{U_S}\mid=5 $). From Fig. \ref{vnfs_vs_links}, it can be observed that when the number of VNFs  of SFCs increases since more resources should be allocated to them,  the objective function of problem \eqref{nfv-problem} increases. Moreover, as physical link bandwidth increases, more bandwidth of lower cost links can be allocated to SFCs. Therefore,  the cost of utilized physical link bandwidth reduced which results in decreasing the objective function of  problem \eqref{nfv-problem}.
	 
	\subsection{Comparison of Our Proposed ARA algorithm with the Proposed Algorithm in \cite{rw-nfv-5}}
	  In Figs. \ref{delay_vs_servers_comparison}, \ref{energy_vs_sfcs_comparison}, and \ref{delay_vs_vnfs_comparison}, we compare the performance of  ARA algorithm  with the HCA algorithm proposed  in \cite{rw-nfv-5}. For a fair comparison, to implement  the ARA algorithm we omit the constraint $ \mathrm{C2} $ \eqref{node-allocation-different server} in problem \eqref{nfv-problem}.    
		
	Fig. \ref{delay_vs_servers_comparison} illustrates the number of active servers versus the number of SFCs and processing capacity of the servers. As can be seen from Fig. \ref{delay_vs_servers_comparison} with increasing the number of SFCs, the number of active servers for providing services to these SFCs is increased.  Also, the number of active servers can be reduced by increasing the processing capacity of the servers due to the implementation of more VNFs on each server. Moreover, Fig. \ref{delay_vs_servers_comparison} shows that the number of active servers in the ARA algorithm is lower than the HCA algorithm proposed in \cite{rw-nfv-5}.	
	
	  In Fig.  \ref{energy_vs_sfcs_comparison}, we compare the energy consumption of the ARA algorithm with the HCA algorithm in \cite{rw-nfv-5}. As observed from Fig.  \ref{energy_vs_sfcs_comparison}, with increasing the number of SFCs, energy consumption is increased.  Also, energy consumption is reduced when the processing capacity of servers is increased. Furthermore,  Fig.  \ref{energy_vs_sfcs_comparison} illustrates that the ARA algorithm outperforms the HCA algorithm proposed in \cite{rw-nfv-5} in terms of energy consumption.
	
	Fig.  \ref{delay_vs_vnfs_comparison} illustrates the average delay of SFCs versus the number of VNFs. From Fig.  \ref{delay_vs_vnfs_comparison}, it is observed that when the number of VNFs of each SFC is increased since more VNFs should be implemented to complete each SFC, the average delay is increased.  Besides, increasing the bandwidth of physical links decreases the average delay because the transmission delay over links is reduced.	  
	 
	\subsection{Performance of the Solution of MO Problem \eqref{problem-mining} }
	\begin{figure}
		\begin{subfigure}{.5\textwidth}
			\centering
			% include first image
			\includegraphics[width=.8\linewidth]{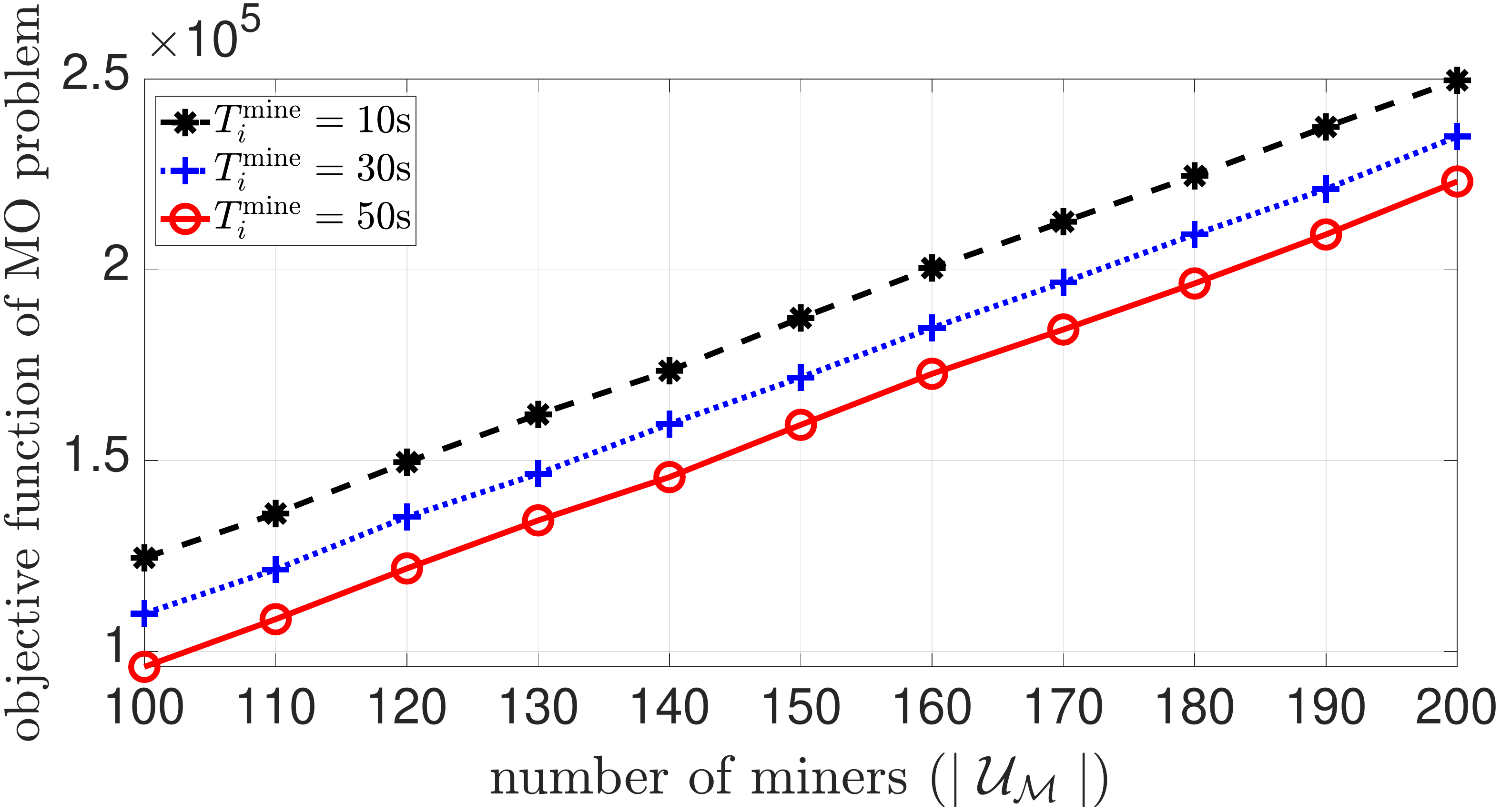}  
			\caption{}
			\label{mining_a}
		\end{subfigure}
		\begin{subfigure}{.5\textwidth}
			\centering
			% include second image
			\includegraphics[width=.8\linewidth]{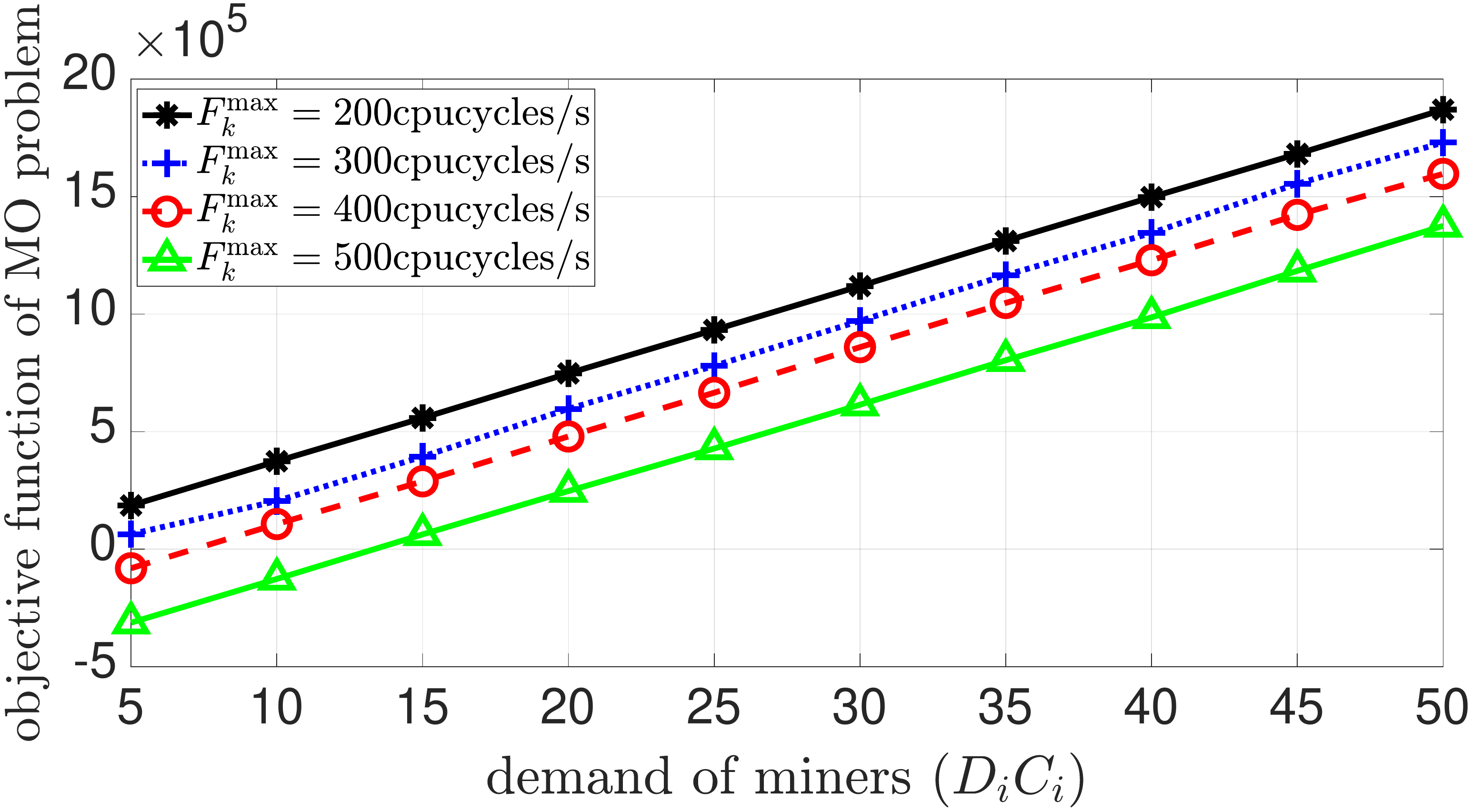}  
			\caption{}
			\label{mining_b}
		\end{subfigure}
		\caption{\small Objective function of MO problem \eqref{problem-mining} vs. \small (a)  maximum tolerable delay	 and number of miners and (b) miners' demand and processing capacity of users}
		\label{delay_vs_miners_number}
		\vspace{-0.8 em}
	\end{figure}

	In this section, we simulate the optimal solution to address  problem \eqref{problem-mining}.
	To do so, similar to \cite{pathgain}, the path gain from each miner to users participating in mining process is modeled by $h_{i,k}=\nu d_{i,k}^{-\mu} $, where $ d_{i,k} $ is the distance between miner $ i $ and user $ k $ which is randomly set, $ \nu $ is a random value that is generated by the Rayleigh distribution, and $ \mu=3 $ is the path loss exponent.
	The other simulation parameters are given in Table \ref{simulation-parameter-mining}. 
	
	\begin{table}
		\centering
		\caption{simulation parameters for mining process } \label{simulation-parameter-mining}
		
		\begin{tabular}{| p{4cm} |p{3.5cm}| }
			\hline\xrowht[()]{5 pt}
			\textbf{Parameter} & \textbf{Value}  \\  % inserts table
			%heading
			\hline 
			\hline\xrowht[()]{5 pt}
			$ \gamma $ & 0.5\\
			\hline\xrowht[()]{5 pt}
			number of users participating in mining process ($ \mid\mathcal{K}_i\mid $) & $ 5 $\\
			\hline\xrowht[()]{5 pt}
			noise power ($ \sigma_k $) & $ 10^{-14} $ \\
			\hline\xrowht[()]{5 pt}
			unit price of each  CPU cycle   ($ \mathrm{cost}_{i,k} $)  &  random selection from $ [1,10] $ \\
			\hline\xrowht[()]{5 pt}			
			users' processing capacity ($F_k^{\mathrm{max}} $)	 &  random selection of $[100,500]~\mathrm{cpucycle/s}$ \\		
			\hline\xrowht[()]{5 pt}
			power consumption 	of user $ k $ ($ \widetilde{p}_k$)		      &    random selection $[0.1,0.9]~\mathrm{W}$ \\
			\hline\xrowht[()]{5 pt}
			transmit power 	of miner $ i $ ($ {p}_{i,k}$)		      &    random selection $[1,10]~\mathrm{mW}$ \\
			\hline\xrowht[()]{5 pt}
			number of transactions in each block  ($ N_{\mathrm{Trans}} $) & 5\\
			\hline\xrowht[()]{5 pt}
			constant reward ($ R_{\mathrm{const}} $)  &   12.5\\
			\hline\xrowht[()]{5 pt}
			variable reward for each transaction ($ R_{\mathrm{Trans}} $) & 0.01\\
			\hline\xrowht[()]{5 pt}
			network latency parameter ($ z $) & 0.01\\
			\hline
		\end{tabular}
	\end{table}
	
	Fig. \ref{mining_a} shows the objective function of MO problem \eqref{problem-mining} with respect to the miner users' maximum tolerable delay  and the number of miner users. It can be seen, with decrease of the miner users' maximum tolerable delay, since the mining process needs to be offloaded to more users and the increasing number of users participating in the mining increases the cost of the miner users paying them, the objective function of  problem \eqref{problem-mining} increases. On the other hand, when the number of miner users increase, because all these miner users want to perform the mining process in order to receive rewards, both the energy consumption and the cost increase resulting in increasing the objective function of \eqref{problem-mining}. Also, in Fig. \ref{mining_b}  the objective function of  MO \eqref{problem-mining}  versus the miner users' demand and the processing capacity of the participating users is shown. From Fig. \ref{mining_b}, it can be observed that by increasing the processing capacity of users participating in the mining process, because the miner users can offload the mining process to fewer number of users, the miner user's energy consumption for sending the mining task to users is reduced. In addition, the miner user can offload most of the mining process to the less costly users. Therefore, by increasing the processing capacity of participating users in the mining process, the objective function of MO problem \eqref{problem-mining} is reduced. On the other hand, as miner users' demand increases, since the miner users have to offload the mining process to more users, both energy consumption and payment costs increase, leading to an increase in the objective function of MO problem \eqref{problem-mining}.

	\section{Conclusion}\label{conclusion}
	In this paper, we studied the NFV resource allocation. To do so, we formulated the NFV-RA problem aimed at minimizing the energy consumption and cost of utilized resources as a multi-objective problem. Due to the NP-hardness of this problem, we proposed two near-optimal algorithms named ARA and HuRA. For the ARA algorithm, we converted the binary variables into continuous ones by adding a penalty function to the objective function. Furthermore, to convexify NFV-RA problem, we employed Majorization-Minimization approximation method. By doing so, the transformed problem becomes a convex problem that can be solved by optimization software packages. In addition, to reduce the complexity of the ARA algorithm, we proposed the HuRA algorithm. For the HuRA algorithm, we decomposed the main NFV-RA problem into two sub-problems namely VNF placement and routing sub-problems. The VNF placement sub-problem is an ILP problem that is NP-hard. To address the VNF placement sub-problem, we employed the Hungarian assignment algorithm. The routing sub-problem is also an LP problem that can be optimally solved in polynomial time. The simulation results illustrated that our proposed ARA and HuRA algorithms achieve near-optimal performance with lower computational complexity. Also, the ARA algorithm outperformed the existing algorithms in terms of the number of active servers, energy consumption, and average delay. 
	On the other hand, resource allocation in  NFV is done by  NFVO as a centralized authority. The NFVO has drawbacks such as a single point of failure and security issues. Relying on the mining process, blockchain technology can address these pitfalls. In wireless networks, the mining process is performed by mobile users who have limited processing resources and cannot perform the mining process lonely. Hence, we assume that miner users can perform the mining process with the participation of other users. The mining process has a lot of energy consumption. Additionally, miner users have to pay to other users for using their processing resources. Therefore, we formulated the problem of minimizing energy consumption and cost for miner users as an LP problem that can be optimally solved in polynomial time. 
	  As future work, one may consider queue latency on servers, which is more practical. In data centers, each server has a queue where users wait to receive processing resources. Considering this causes that the latency model will be different from that of our paper, so, our proposed algorithms should be accordingly revised to take the queuing latency into account.  Furthermore,   it can be assumed that users are with limited energy, and energy is harvested. To do so, an energy constraint is added to the constraints of the optimization problem, according to which the total energy consumption of users should not be larger than the total initial energy and harvested energy. By adding this constraint to the optimization problem, the solution proposed in this paper should be modified to ensure this constraint.

\begin{IEEEbiography}
[{\includegraphics[width=1in,height=1.25in,clip,keepaspectratio]{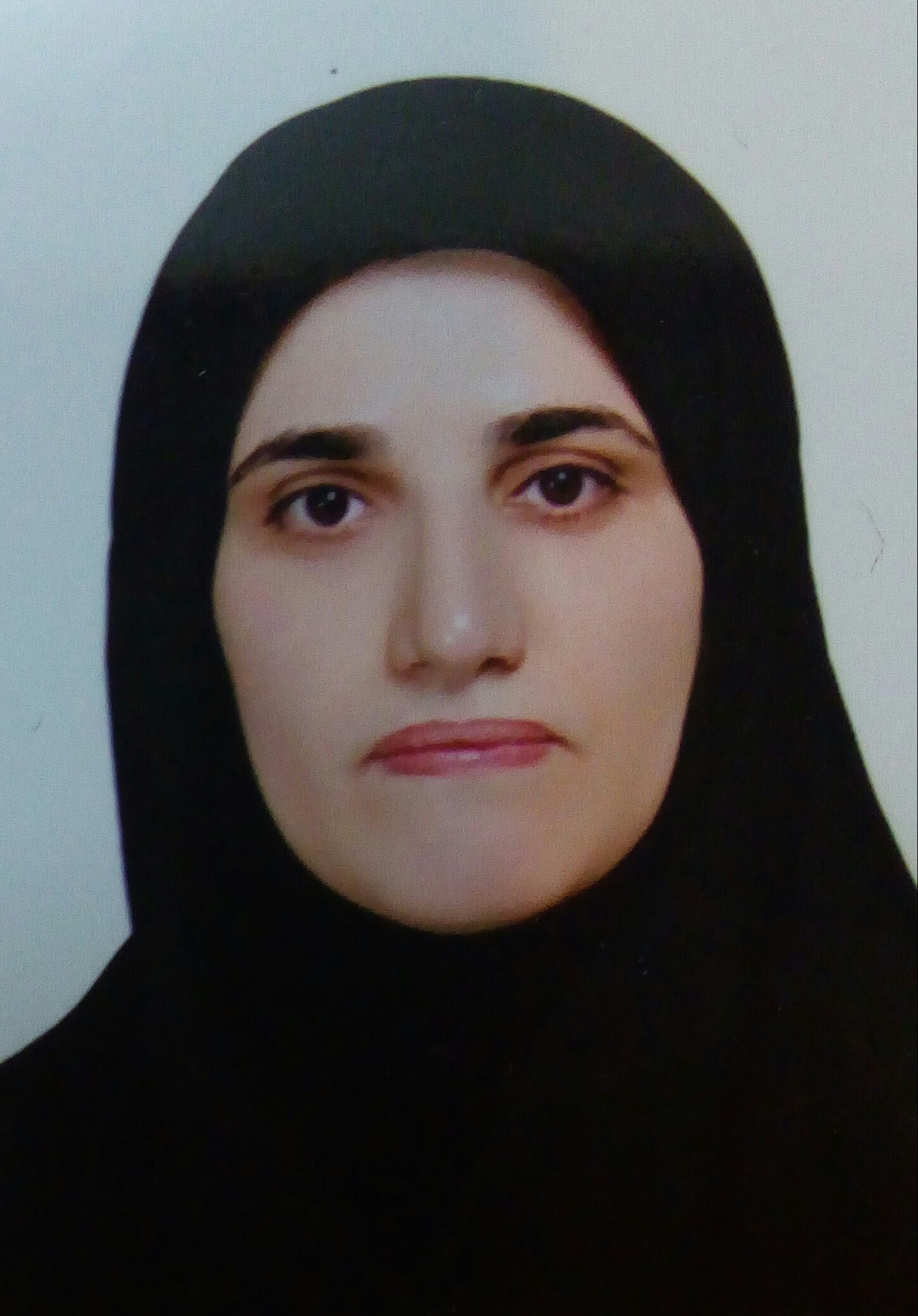}}]{Shiva Kazemi Taskou}
is a PhD candidate at the Department of Computer Engineering, Amirkabir University
of Technology, Tehran, Iran. She received her  M.Sc. degree from  Amirkabir University of Technology, and B.Sc. degree from Payame Nour University, Tehran, Iran, in 2017 and 2014, respectively. Her current research interests include resource management in wireless networks, wireless network virtualization, and optimization.
\end{IEEEbiography}

\begin{IEEEbiography}[{\includegraphics[width=1in,height=1.25in,clip,keepaspectratio]{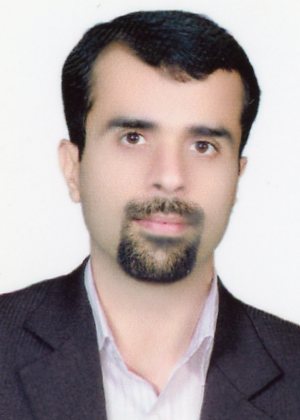}}]{Mehdi Rasti}
	[S'08-M'11] 
	received the B.Sc. 	degree in electrical engineering from Shiraz
	University, Shiraz, Iran, in 2001, and the M.Sc. and 	Ph.D. degrees in electrical engineering from Tarbiat Modares University, Tehran, Iran, in 2003 and 2009, respectively. He is currently an Assistant Professor 	with the Department of Computer Engineering, Amirkabir University of Technology, Tehran, Iran. From 	November 2007 to November 2008, he was a Visiting Researcher with the Wireless@KTH, Royal Institute of Technology, Stockholm, Sweden. From 	September 2010 to July 2012, he was with the Shiraz University of Technology, Shiraz. From June 2013 to August 2013, and from July 2014 to 	August 2014, he was a Visiting Researcher with the Department of Electrical 	and Computer Engineering, University of Manitoba, Winnipeg, MB, Canada. 	His current research interests include radio resource allocation in IoT, beyond	5G, and 6G wireless networks.
\end{IEEEbiography}

\begin{IEEEbiography}[{\includegraphics[width=1in,height=1.25in,clip,keepaspectratio]{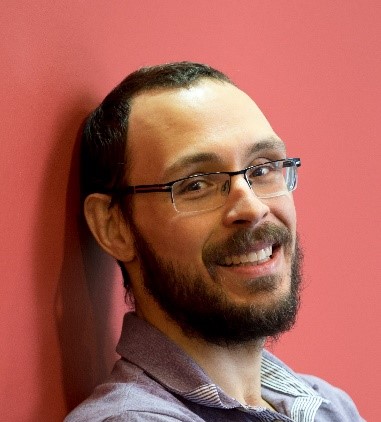}}]{Pedro H. J. Nardelli} [M'07, SM'19]
	received the B.S. and M.Sc. degrees in electrical engineering from the State University of Campinas, Brazil, in 2006 and 2008, respectively. In 2013, he received his doctoral degree from University of Oulu, Finland, and State University of Campinas following a dual degree agreement. He is currently Associate Professor (tenure track) in IoT in Energy Systems at LUT University, Finland, and holds a position of Academy of Finland Research Fellow with a project called Building the Energy Internet as a large-scale IoT-based cyber-physical system that manages the energy inventory of distribution grids as discretized packets via machine-type communications (EnergyNet). He leads the \href{https://cps-g.com/}{Cyber-Physical Systems Group}  at LUT, and is Project Coordinator of the CHIST-ERA European consortium Framework for the Identification of Rare Events via Machine Learning and IoT Networks (FIREMAN) and of the project Swarming Technology for Reliable and Energy-aware Aerial Missions (STREAM) supported by Jane and Aatos Erkko Foundation. He is also Docent at University of Oulu in the topic of \enquote{communications strategies and information processing in energy systems}. His research focuses on wireless communications particularly applied in industrial automation and energy systems. He received a best paper award of IEEE PES Innovative Smart Grid Technologies Latin America 2019 in the track \enquote{Big Data and Internet of Things}. He is also IEEE Senior Member. More information: \url{https://sites.google.com/view/nardelli/}
	 
\end{IEEEbiography}
\end{document}

	%%%%%%%%%%%%%%%%%%%%%%%%%%%%%%%%%%%%%%%%%%%%%%%%%%%%%%
	\newcommand{\labelBlock}[1]{%
		\refstepcounter{BlockCounter}%
		\hypertarget{#1}{}(\theBlockCounter\label{#1})%
	}
	
	\newcommand{\refBlock}[1]{%
		\hyperref[#1]{Block~\ref*{#1}}% (see Problem 18 of the hyperref manual)
	}
	\newcommand*\colvec[1]{\begin{pmatrix}#1\end{pmatrix}}
	\makeatletter
	\newcommand*{\rom}[1]{\expandafter\@slowromancap\romannumeral #1@}
	\makeatother
	
	\title{Energy and Cost Efficient Resource Allocation for Blockchain-Enabled NFV}% ,~\IEEEmembership{Student Member,~IEEE}  ,
	\author{Shiva Kazemi Taskou  and Mehdi Rasti,~\IEEEmembership{Member,~IEEE} 
		\thanks{S. Kazemi Taskou and M. Rasti are with Department of Computer Engineering, Amirkabir University of Technology, Tehran, Iran. (e-mail: \{shiva.kt,  rasti\}@aut.ac.ir)}
	}
	%\maketitle
	
	%\markboth{Copyright (c) 2017 IEEE. Personal use of this material is permitted. However, permission to use this material for any other purposes must be obtained from the IEEE by sending a request to pubspermissions@ieee.org.}	{Kazemi Taskou \MakeLowercase{\textit{et al.}}: Fast Water-Filling Method for Sum-Power Minimization in OFDMA  Networks}
\section*{Appendix A. Table of Notations}
%\subsection*{A. Table of Notations}
The list of notations for NFV resource allocation and blockchain networks are represented in Table \ref{notation-nfv} and Table \ref{notation-bc}, respectively.  
\begin{table}
	\renewcommand\thetable{ 4}
	\centering
	\caption{  List of Network Function Virtualization Notations  } \label{notation-nfv}
	
	\begin{tabular}{ |p{4cm}|| p{4cm} |}
		\hline \xrowht[()]{5 pt} %inserts double horizontal lines
		\textbf{Notation} & \textbf{Description}  \\  % inserts table
		\hline
		\hline \xrowht[()]{5 pt} 
		$ \mathcal{U_S} $ & the set of SFC requesting users\\
		\hline \xrowht[()]{5 pt}
		$ \mathcal{G} = (\mathcal{V},\mathcal{L}) $ & directed graph\\
		\hline \xrowht[()]{5 pt}
		$\mathcal{V}=\{\mathcal{AC},\mathcal{TR},\mathcal{N}\} $  &  the set of servers \\
		\hline \xrowht[()]{5 pt}
		$ \mathcal{L} $ & the set of directed links \\
		\hline \xrowht[()]{5 pt}
		$ \mathcal{AC} $ & the access switches (source nodes)\\
		\hline \xrowht[()]{5 pt}
		$ \mathcal{TR} $ & the transport switches (destination nodes)\\
		\hline \xrowht[()]{5 pt}
		$ \mathcal{N} $ &  the processing servers\\
		\hline \xrowht[()]{5 pt}
		$ C_n^{\mathrm{max}}$  &  maximum processing capacity of server $ n $ \\
		\hline \xrowht[()]{5 pt}
		$ B_l^{\mathrm{max}} $   & maximum bandwidth of link $ l $ \\
		\hline \xrowht[()]{5 pt}
		$ \mathcal{S}_i $ & the SFC for user $ i $\\
		\hline \xrowht[()]{5 pt}
		$\mathcal{S}_i[1]  $ & AMF \\
		\hline \xrowht[()]{5 pt}
		$ \mathcal{S}_i[J_i] $ & UPF\\
		\hline \xrowht[()]{5 pt}
		$ \mathcal{S}_i[2], \mathcal{S}_i[3],\cdots , \mathcal{S}_i[J_i-1] $ & SMF and other VNFs \\
		\hline \xrowht[()]{5 pt}
		$ \mathcal{S}_i[0] $  &  source node\\
		\hline \xrowht[()]{5 pt}
		$ \mathcal{S}_i[J_i +1] $& destination node\\
		\hline \xrowht[()]{5 pt}
		$ C_{i,j} $  &  required CPU cycles for each VNF $ j $ of SFC $ \mathcal{S}_i $ \\
		\hline \xrowht[()]{5 pt}
		$ B^{j,j+1}_i $  &  the required bandwidth for links between $ j $th and $ j+1 $th VNFs of $ \mathcal{S}_i $\\
		\hline \xrowht[()]{5 pt}
		$ T_i^{\mathrm{th}} $  &  end-to-end  maximum tolerable delay \\
		\hline \xrowht[()]{5 pt}
		$ T_i $  &  The end-to-end delay of SFC $ \mathcal{S}_i $\\
		\hline \xrowht[()]{5 pt}
		$ p_n^s $ & static power to be active \\
		\hline \xrowht[()]{5 pt}
		$ p_n $ & consumption power of processing at each second\\
		\hline \xrowht[()]{5 pt}
		$ E $ &  energy consumed in the data center\\
		\hline \xrowht[()]{5 pt}
		$ \mathrm{cost}_{i,n} $ & the unit price of each  CPU cycle of server $ n $\\
		\hline \xrowht[()]{5 pt} 
		$ \mathrm{cost}_{i,l} $  &  the unit price for transmitting  each bit per second  over physical link $ l $\\
		\hline \xrowht[()]{5 pt}
		$ \mathrm{cost} $ &  the cost of resources  used by  all  SFC requesting users \\
		\hline \xrowht[()]{5 pt}
		$ \alpha $ & a weighted factor\\
		\hline
		\hline \xrowht[()]{5 pt}
		\textbf{Variables} & \textbf{Description} \\
		\hline
		\hline \xrowht[()]{5 pt}
		$ x_{i,j}^n $ & embedding of $ j $th VNF of user $ i $'s SFC on server $ n $\\
		\hline \xrowht[()]{5 pt}
		$ y_{l_j^{j+1}}^i $ &  bandwidth allocation of physical link $ l $ to virtual link between $ j $th and $ j + 1 $th VNFs of user $ i $'s SFC\\ 
		\hline \xrowht[()]{5 pt}
		$ \beta_n $ & servers status (active or inactive)\\	
		
		\hline
	\end{tabular}
\end{table}

\begin{table}
	\renewcommand\thetable{5}
	\centering
	\caption{  List of Blockchain Network Notations  } \label{notation-bc}
	
	\begin{tabular}{ |p{4cm}|| p{4cm} |}
		\hline \xrowht[()]{5 pt} %inserts double horizontal lines
		\textbf{Notation} & \textbf{Description}  \\ 
		\hline
		\hline \xrowht[()]{5 pt}
		$ \mathcal{U_M} $ & the set of miner users \\
		\hline \xrowht[()]{5 pt}
		$ D_i $ & size of mining task in bits\\
		\hline \xrowht[()]{5 pt}
		$ C_i $ & required CPU cycles of each bit of mining task\\
		\hline \xrowht[()]{5 pt}
		$ \mathcal{K}_i $   &    a set of users participating in mining process\\
		\hline \xrowht[()]{5 pt}
		$ \mathrm{cost}_{i,k} $  & paying cost to the user $ k $ for each CPU cycle\\
		\hline \xrowht[()]{5 pt}
		$ T_{i,k}^{tr} $  &  transmission time for sending a portion of mining task to user $ k $\\
		\hline \xrowht[()]{5 pt}
		$ R_{i,k} $ &  the data rate between  miner user $ i $ and user $ k $\\
		\hline \xrowht[()]{5 pt}
		$ p_{i,k} $ & the transmit power of miner user $ i $ to user $ k $\\
		\hline \xrowht[()]{5 pt}
		$ h_{i,k} $ & the path-gain from miner user $ i $ toward user $ k $\\
		\hline \xrowht[()]{5 pt}
		$ \sigma_k $ & the noise power at user $ k $ \\
		\hline \xrowht[()]{5 pt}
		$ T_{i,k}^{proc}$    &   processing time \\
		\hline \xrowht[()]{5 pt} 
		$ F_k^{\mathrm{max}} $ & the maximum CPU cycles per second of user $ k $'s device\\
		\hline \xrowht[()]{5 pt}
		$ \widetilde{p}_k $ & power consumption of user $ k $'s device at each second\\
		\hline \xrowht[()]{5 pt}
		$ E_{\mathrm{mine}} $ & the total energy consumption during mining process \\
		\hline \xrowht[()]{5 pt}
		$ R_{\mathrm{Trans}} $ & reward of each transaction\\
		\hline \xrowht[()]{5 pt}
		$ R_{\mathrm{const}} $ &  a constant  reward value \\
		\hline \xrowht[()]{5 pt}  
		$ N_{Trans} $ & the number of transactions in the mined block\\
		\hline \xrowht[()]{5 pt}
		$ p_{\mathrm{orphan}}$ & the probability of a block orphaning \\
		\hline \xrowht[()]{5 pt}
		$ \gamma $ &a weighted factor\\
		\hline
		\hline \xrowht[()]{5 pt}
		\textbf{Variables} & \textbf{Description} \\
		\hline
		\hline \xrowht[()]{5 pt}		
		$ f_{i,k} $ &  the weight of mining task $ i $ allocated to user $ k\in \mathcal{K}_i $ \\
		\hline
	\end{tabular}
\end{table}

\section*{Appendix B. Proof of Theorem 1}
To prove Theorem 1, let us define Lagrangian multipliers $ \lambda_1 $ and $ \lambda_2 $ to handle non-convexity of constraints $\mathrm{C8.1}$ and $\mathrm{C9.1}$ in (19) and (20), respectively. Using the abstract Lagrangian duality [43], we define $ \mathcal{L}(\boldsymbol{\beta,X,Y},\lambda_1,\lambda_2)$ as 
\begin{equation}\label{abstract Lagrangian duality}
\begin{aligned}
&\mathcal{L}(\boldsymbol{\beta,X,Y},\lambda_1,\lambda_2) =F(\boldsymbol{\beta},\boldsymbol{X},\boldsymbol{Y})+ \lambda_1 \sum\limits_{n\in\mathcal{N}}(\beta_n-\beta_n^2)\\
&+\lambda_2 \sum\limits_{n\in\mathcal{N}}\sum\limits_{i\in\mathcal{U_S}}\sum\limits_{j\in\mathcal{S}_i}(x_n^{i,j}-{x_n^{i,j}}^2). 
\end{aligned}
\end{equation}
By doing so,  problem (21) can be  expressed as
\begin{equation}\label{rewrite-problem-17}
\begin{aligned}
\displaystyle \min_{\substack{\boldsymbol{\beta,X,Y}}}\displaystyle \max_{\substack{\lambda_1,\lambda_2\geq 0}}\mathcal{L}(\boldsymbol{\beta,X,Y},\lambda_1,\lambda_2).
\end{aligned}
\end{equation} 
Furthermore, the dual problem of (21) is
\begin{equation}\label{dual}
\begin{aligned}
\displaystyle \max_{\substack{\lambda_1,\lambda_2\geq 0}}\min_{\substack{\boldsymbol{\beta,X,Y}}}\displaystyle \mathcal{L}(\boldsymbol{\beta,X,Y},\lambda_1,\lambda_2)=\displaystyle\max_{\substack{\lambda_1,\lambda_2\geq 0}} \Theta(\lambda_1,\lambda_2),
\end{aligned}
\end{equation}
where $ \Theta(\lambda_1,\lambda_2)=\displaystyle \min_{\substack{\boldsymbol{\beta,X,Y}}}\displaystyle \mathcal{L}(\boldsymbol{\beta,X,Y},\lambda_1,\lambda_2) $. 

By introducing  $ d^* $ as  the optimal  solution of problem (14),	according to the weak duality, there is a duality gap, i.e.,
\begin{equation}\label{gap}
\begin{aligned}	
&\max_{\substack{\lambda_1,\lambda_2\geq 0}} \Theta(\lambda_1,\lambda_2)=\max_{\substack{\lambda_1,\lambda_2\geq 0}}\min_{\substack{\boldsymbol{\beta,X,Y}}}\displaystyle \mathcal{L}(\boldsymbol{\beta,X,Y},\lambda_1,\lambda_2)
\\&\leq 	\displaystyle \min_{\substack{\boldsymbol{\beta,X,Y}}}\displaystyle \max_{\substack{\lambda_1,\lambda_2\geq 0}}\mathcal{L}(\boldsymbol{\beta,X,Y},\lambda_1,\lambda_2)=d^*
\end{aligned}
\end{equation}

Since   $ \sum\limits_{n\in\mathcal{N}}(\beta_n-{\beta_n}^2)\geq 0 $ and  $ \sum\limits_{n\in\mathcal{N}}\sum\limits_{i\in\mathcal{U_S}}\sum\limits_{j\in\mathcal{S}_i}(x_n^{i,j}-{x_n^{i,j}}^2)\geq 0 $,  $ \mathcal{L}(\boldsymbol{\beta,X,Y},\lambda_1,\lambda_2) $  is increasing in $ \lambda_1$ and $\lambda_2 $. This means that $\Theta(\lambda_1,\lambda_2)$ is increasing in $ \lambda_1$ and $\lambda_2 $ and bounded by the optimal solution of problem (21). Regarding the solution of \eqref{dual} (dual problem of (21)), two cases are considered.

\begin{enumerate}
	\item For the first case, we assume that $ \sum\limits_{n\in\mathcal{N}}(\beta_n-{\beta_n}^2)= 0 $ and  $ \sum\limits_{n\in\mathcal{N}}\sum\limits_{i\in\mathcal{U_S}}\sum\limits_{j\in\mathcal{S}_i}(x_n^{i,j}-{x_n^{i,j}}^2)= 0 $, so the optimal solution of dual problem of (21), i.e., $ \widetilde{\lambda_1} $, $ \widetilde{\lambda_2} $, $ \widetilde{\boldsymbol{\beta}} $, $ \widetilde{\boldsymbol{X}} $, and $ \widetilde{\boldsymbol{Y}} $ is a feasible solution to problem (21). Therefore, by substituting  $ \widetilde{\boldsymbol{\beta}} $, $ \widetilde{\boldsymbol{X}} $, and $ \widetilde{\boldsymbol{Y}} $ in problem (14), we have
	
	\begin{equation}\label{case-1}
	\begin{aligned}
	\Theta(\widetilde{{\lambda_1}},\widetilde{{\lambda_2}})=\mathcal{L}(\widetilde{\boldsymbol{\beta}},\widetilde{\boldsymbol{X}},\widetilde{\boldsymbol{Y}},\widetilde{{\lambda_1}},\widetilde{{\lambda_2}})=F(\widetilde{\boldsymbol{\beta}},\widetilde{\boldsymbol{X}},\widetilde{\boldsymbol{Y}})\geq d^*.
	\end{aligned}
	\end{equation}
	According to \eqref{gap} and \eqref{case-1}, we can conclude that there is not gap between  problem (14)  and the dual problem of (21), i.e.,
	\begin{equation}\label{gap-zero}
	\begin{aligned}	
	&\max_{\substack{\lambda_1,\lambda_2\geq 0}}\min_{\substack{\boldsymbol{\beta,X,Y}}}\displaystyle \mathcal{L}(\boldsymbol{\beta,X,Y},\lambda_1,\lambda_2)\\
	&=	\displaystyle \min_{\substack{\boldsymbol{\beta,X,Y}}}\displaystyle \max_{\substack{\lambda_1,\lambda_2\geq 0}}\mathcal{L}(\boldsymbol{\beta,X,Y},\lambda_1,\lambda_2).
	\end{aligned}
	\end{equation}
	Moreover, since $ \Theta({\lambda_1,\lambda_2}) $ is monotonic with respect to $ \lambda_1 $ and $ \lambda_2 $, we have $ \Theta({\lambda_1,\lambda_2})=d^*, ~\forall \lambda_1 \geq \widetilde{{\lambda_1}} $ and $ ~\forall \lambda_2\geq \widetilde{{\lambda_2}}  $. This means that the optimal solution of the dual problem of (21) is equal to the optimal solution of problem (14).
	
	\item For the second case, we assume that $ \sum\limits_{n\in\mathcal{N}}(\beta_n-{\beta_n}^2)> 0 $ and $ \sum\limits_{n\in\mathcal{N}}\sum\limits_{i\in\mathcal{U_S}}\sum\limits_{j\in\mathcal{S}_i}(x_n^{i,j}-{x_n^{i,j}}^2)> 0 $. In this case, due to the monotonicity of $ \Theta({\lambda_1,\lambda_2}) $ with respect to $ \lambda_1 $ and $ \lambda_2 $, the dual problem of (21) is unbounded, i.e., $  \Theta(\widetilde{\lambda_1},\widetilde{\lambda_2})\rightarrow \infty $. Since, the primal problem  has a finite solution, according to \eqref{gap}, the dual problem cannot be infinite. In the other words, the infinity of the dual problem of (21) is in conflict with \eqref{gap}. So, for the optimal solution, we have $ \sum\limits_{n\in\mathcal{N}}(\beta_n-{\beta_n}^2)= 0 $ and $ \sum\limits_{n\in\mathcal{N}}\sum\limits_{i\in\mathcal{U_S}}\sum\limits_{j\in\mathcal{S}_i}(x_n^{i,j}-{x_n^{i,j}}^2)= 0 $  which completes the proof.
\end{enumerate}

\subsection*{Appendix C. Proof of Proposition 1}
To prove Proposition 1, we denote the objective function of problem (22) at iteration $ t $ as $ f(\boldsymbol{\beta}^t,\boldsymbol{X}^t,\boldsymbol{Y}^t)-g(\boldsymbol{\beta}^t,\boldsymbol{X}^t) $. Since to approximate $ \sum\limits_{n\in\mathcal{N}}{\beta_n}^2 $ and $ \sum\limits_{n\in\mathcal{N}}\sum\limits_{i\in\mathcal{U_S}}\sum\limits_{j\in\mathcal{S}_i}{x_n^{i,j}}^2 $, we employ  first-order Taylor approximation method and from this fact that the first-order Taylor approximation is a lower bound for convex functions [44], we have
\begin{equation}\label{proof}
\begin{aligned}
&f(\boldsymbol{\beta}^{t+1},\boldsymbol{X}^{t+1},\boldsymbol{Y}^{t+1})-g(\boldsymbol{\beta}^{t+1},\boldsymbol{X}^{t+1})\\
&\leq f(\boldsymbol{\beta}^{t+1},\boldsymbol{X}^{t+1},\boldsymbol{Y}^{t+1})-g(\boldsymbol{\beta}^{t},{\boldsymbol{X}^{t}})\\&+\nabla_ {\boldsymbol{\beta}}g(\boldsymbol{\beta}^{t},{\boldsymbol{X}^{t+1}})(\boldsymbol{\beta}^{t+1}-{\boldsymbol{\beta}^{t}})\\&+\nabla_ {\boldsymbol{X}}g(\boldsymbol{\beta}^{t+1},{\boldsymbol{X}^{t}})(\boldsymbol{X}^{t+1}-{\boldsymbol{X}^{t}})\\
&\leq f(\boldsymbol{\beta}^{t},\boldsymbol{X}^{t},\boldsymbol{Y}^{t})-g(\boldsymbol{\beta}^{t},{\boldsymbol{X}^{t}})+\nabla_ {\boldsymbol{\beta}}g(\boldsymbol{\beta}^{t},{\boldsymbol{X}^{t}})(\boldsymbol{\beta}^t-{\boldsymbol{\beta}^{t}})\\&+\nabla_ {\boldsymbol{X}}g(\boldsymbol{\beta}^{t},{\boldsymbol{X}^{t}})(\boldsymbol{X}^t-{\boldsymbol{X}^{t}})=f(\boldsymbol{\beta}^{t},\boldsymbol{X}^{t},\boldsymbol{Y}^{t})-g(\boldsymbol{\beta}^{t},{\boldsymbol{X}^{t}}).
\end{aligned}
\end{equation}
The equation \eqref{proof} means that the optimal solution of problem (22) at each iteration $ t $ is better than previous iteration $ t-1 $. So, obtaining the optimal solution of problem (22) iteratively continues until $\boldsymbol{\beta}^{t}=\boldsymbol{\beta}^{t-1}  $, $ \boldsymbol{X}^{t}=\boldsymbol{X}^{t-1} $, and $ \boldsymbol{Y}^{t}=\boldsymbol{Y}^{t-1} $. 

Furthermore, when $\boldsymbol{\beta}^{t}=\boldsymbol{\beta}^{t-1}  $, $ \boldsymbol{X}^{t}=\boldsymbol{X}^{t-1} $, and $ \boldsymbol{Y}^{t}=\boldsymbol{Y}^{t-1} $, from \eqref{proof}, we have $ f(\boldsymbol{\beta}^{t+1},\boldsymbol{X}^{t+1},\boldsymbol{Y}^{t+1})-g(\boldsymbol{\beta}^{t+1},\boldsymbol{X}^{t+1})= f(\boldsymbol{\beta}^{t},\boldsymbol{X}^{t},\boldsymbol{Y}^{t})-g(\boldsymbol{\beta}^{t},{\boldsymbol{X}^{t}})$ which proves tightness of the upper bound.\\
Accordingly, the optimal solution of problem (22) is a local optimal solution for problem (21). Based on Theorem 1, problem (21) is equivalent to problem (14) for a sufficiently large $ \lambda_1 \gg 1 $ and $ \lambda_2 \gg 1 $. So,  the optimal solution of problem (22) is a local optimal solution for problem (14).

\section*{Appendix D. Pseudo-code of Algorithm 2- Our Proposed HuRA Algorithm}
\renewcommand\thealgocf{ 2}
\begin{algorithm*}[ht]\label{HuRa}

	\BlankLine

	\SetKwFunction{Range}{range}%%
	\SetKw{KwTo}{in}\SetKwFor{For}{for}{\string:}{}%
	\SetKwIF{If}{ElseIf}{Else}{if}{:}{elif}{else:}{}%
	
	\SetAlgoNoEnd
	
	\SetAlgoNoLine%
	\SetKwInOut{Input}{Input}
	\SetKwInOut{Output}{Output}
	Sort the SFCs by increasing the maximum tolerable delay\\
	Initialize $ \mathrm{ remainC}_n=C_n^{\mathrm{max}}$ %,~\forall n \in\mathcal{N}$ 
	and $ \mathrm{ remainB}_l=B_l^{\mathrm{max}} $,%,~\forall l \in\mathcal{L}$,
	~$ \mathrm{objective}=0 $,  $ \mathrm{ candidate}_{i,j}=[] $, $ \beta_n=0 $.\\
	\textbf{for} each SFC $ \mathcal{S}_i $, $ i\in\mathcal{U_S} $:\\
	\Indp
	\textbf{for} each VNF $ j\in\mathcal{S}_i $:\\
	\Indp
	\textbf{for} each server $ n\in\mathcal{N} $:\\
	\Indp
	\textbf{if} $ \mathrm{ remainC}_n \geq C_{i,j}$:\\
	\Indp
	$  \mathrm{ candidate}_{i,j}= \mathrm{ candidate}_{i,j}\cup n $.\\
	\Indm
	\text{\textbf{end if}}\\
	\Indm
	\text{\textbf{end for}}\\
	\textbf{for} each server $ n\in\mathrm{ candidate}_{i,j} $:\\
	\Indp
	\textbf{if} $ \beta_n=0 $:\\
	\Indp
	$A[j,n]=\mathrm{objective}+\alpha\big[\beta_np_n^s+p_nC_{i,j}/C_n^{\mathrm{max}}\big]+(1-\alpha)\big[\mathrm{cost}_{i,n}C_{i,j}\big] $\\
	\Indm
	\textbf{else}:\\
	\Indp
	$A[j,n]=\mathrm{objective}+\alpha\big[p_nC_{i,j}/C_n^{\mathrm{max}}\big]+(1-\alpha)\big[\mathrm{cost}_{i,n}C_{i,j}\big] $\\
	\Indm
	\text{\textbf{end if}}\\
	\Indm
	\text{\textbf{end for}}\\
	\Indm
	\text{\textbf{end for}}\\
	\text{Give matrix $ A $ to Hungarian algorithm to obtain matrix $ A' $}\\
	\textbf{if} $ A'[j,n]=1 $:\\
	\Indp
	$ \beta_n=1 $\\
	\Indm
	\text{\textbf{end if}}\\
	\text{Solve  problem (24)}\\
	\textbf{if}	\text{the answer of  problem (24) is feasible}:\\
	\Indp
	\text{algorithm terminates}\\
	\Indm
	\textbf{else}:\\
	\Indp
	\textbf{for} each VNF $ j\in\mathcal{S}_i $:\\
	\Indp
	\textbf{for} each server $ n\in\mathrm{ candidate}_{i,j} $:\\
	\Indp
	$A[j,n]=C_{i,j}/C_n^{\mathrm{max}} $\\
	\Indm
	\text{\textbf{end for}}\\
	\Indm
	\text{\textbf{end for}}\\
	\text{Repeat steps 18-22}\\
	\Indm
	\text{\textbf{end for}}\\
	\text{Update $ \mathrm{ remainC}_n=\mathrm{ remainC}_n-x_n^{i,j}C_{i,j} ,~\forall n \in\mathcal{N}$,} $ \mathrm{ remainB}_l=\mathrm{ remainB}_l-y^i_{l_j^{j+1}} ,~\forall l \in\mathcal{L}$, and $ \mathrm{objective}=\alpha\Big[\sum\limits_{n \in \mathcal{N}}\big( \beta_np_n^s+p_n\sum\limits_{i\in\mathcal{U_S}}\sum\limits_{j \in \mathcal{S}_i}x_n^{i,j}C_{i,j}/C_n^{\mathrm{max}}\big)\Big]+(1-\alpha)\Big[\sum\limits_{n \in \mathcal{N}}\ \mathrm{cost}_{i,n}\sum\limits_{i\in\mathcal{U_S}}\sum\limits_{j \in \mathcal{S}_i}x_n^{i,j}C_{i,j}+ \sum\limits_{l \in \mathcal{L}} \mathrm{cost}_{i,l} \sum\limits_{i\in\mathcal{U_S}} \sum\limits_{j \in \mathcal{S}_i\cup \{0\}}y^i_{l_j^{j+1}} \Big] $
	\caption{Our proposed HuRA algorithm  to solve  problem (14)}	
\end{algorithm*}